\documentclass[twocolumn]{aastex701}

\usepackage{graphicx}	
\usepackage{amsmath}	
\usepackage{tikz}

\newcommand{\planck}{\textit{Planck}}
\newcommand{\co}{$^{12}\rm{CO}$}
\newcommand{\coo}{$^{13}\rm{CO}$}
\newcommand{\cooo}{ $\rm{C}^{18}\rm{O}$}
\newcommand{\cloudname}{G111}
\newcommand{\vrangemin}{[-51.4;~-50.5]}
\newcommand{\vrangemid}{[-52.3;~-51.5]}
\newcommand{\vrangemax}{[-52.8;~-52.3]}
\newcommand{\vrangelast}{[-53.4;~-52.8]}
\newcommand{\bvgt}{B$_{\rm{VGT}}$}
\newcommand{\bplanck}{B$_{\rm{\textit{Planck}}}$}
\newcommand{\bvis}{B$_{\rm{R}}$}
\newcommand{\bnir}{B$_{\rm{H}}$}
\newcommand{\fgd}{$2500$}
\newcommand{\vmin}{$-50.5$}
\newcommand{\vmax}{$-51.4$}
\newcommand{\vminn}{$-51.5$}
\newcommand{\vmaxx}{$-52.3$}
\newcommand{\vminnn}{$-52.8$}
\newcommand{\vmaxxx}{$-53.4$}

\newcommand{\disgaia}{\texttt{r\_med\_photogeo}}
\newcommand{\revision}[1]{{\textcolor{black} {#1}}}
\newcommand{\revisionn}[1]{{\textcolor{black} {#1}}}

\received{2025 February 17}
\revised{YYY}
\accepted{2025 July 2}

\submitjournal{AJ}

\shorttitle{Magnetic Field of \cloudname}
\shortauthors{Alina et al.}

\begin{document}

\title{Magnetic field of a ring-like shape molecular cloud}

\correspondingauthor{Dana Alina}
\email{dana.alina@nu.edu.kz}
\author[0000-0001-5403-356X]{Dana Alina}
\affiliation{Physics Department, School of Sciences and Humanities, Nazarbayev University, Kabanbay batyr ave, 53, 010000 Astana, Kazakhstan}
\email[show]{dana.alina@nu.edu.kz}  

\author{Adel Umirbayeva}
\affiliation{Physics Department, School of Sciences and Humanities, Nazarbayev University, Kabanbay batyr ave, 53, 010000 Astana, Kazakhstan}
\affiliation{Laboratory of Physics of Stars and Nebulae, Fesenkov Astrophysical Institute, Observatory 23, 050020 Almaty, Kazakhstan}
\email{adel.umirbayeva@gmail.com}

\author{Yasuo Doi}
\affiliation{Department of Earth Science and Astronomy, Graduate School of Arts and Sciences, The University of Tokyo, 3-8-1 Komaba, Meguro, Tokyo 153-8902, Japan}
\email{adel.umirbayeva@gmail.com}

\author{Soichiro Jo}
\affiliation{Department of Earth Science and Astronomy, Graduate School of Arts and Sciences, The University of Tokyo, 3-8-1 Komaba, Meguro, Tokyo 153-8902, Japan}
\email{adel.umirbayeva@gmail.com}

\author{Yue Hu}
\affiliation{Institute for Advanced Study, 1 Einstein Drive, Princeton, NJ 08540, USA}
\email{adel.umirbayeva@gmail.com}

\author{Alex Lazarian}
\affiliation{Department of Astronomy, University of Wisconsin-Madison, Madison, WI, 53706, USA}
\email{adel.umirbayeva@gmail.com}

\author{Janik Karoly}
\affiliation{Department of Physics and Astronomy, University College London, WC1E 6BT London, UK}
\email{adel.umirbayeva@gmail.com}

\author{Tie Liu}
\affiliation{Key Laboratory for Research in Galaxies and Cosmology, Shanghai Astronomical Observatory, Chinese Academy of Sciences, 80 Nandan Road, Shanghai 200030, People's Republic of China}
\email{adel.umirbayeva@gmail.com}

\author{Koji S. Kawabata}
\affiliation{Hiroshima Astrophysical Science Center, Hiroshima University, 1-3-1 Kagamiyama, Higashi-Hiroshima, Hiroshima 739-8526, Japan}
\affiliation{Physics Program, Graduate School of Advanced Science and Engineering, Hiroshima University, 1-3-1 Kagamiyama, Higashi-Hiroshima, Hiroshima 739-8526, Japan}
\email{adel.umirbayeva@gmail.com}

\author{Alua Mukhash}
\affiliation{Math Department, School of Sciences and Humanities, Nazarbayev University, Kabanbay batyr ave, 53, 010000 Astana, Kazakhstan}
\email{adel.umirbayeva@gmail.com}

\author{Danial Zhumagayir}
\affiliation{Physics Department, School of Sciences and Humanities, Nazarbayev University, Kabanbay batyr ave, 53, 010000 Astana, Kazakhstan}
\email{adel.umirbayeva@gmail.com}

\author{Tomori Hori}
\affiliation{Physics Program, Graduate School of Advanced Science and Engineering, Hiroshima University, 1-3-1 Kagamiyama, Higashi-Hiroshima, Hiroshima 739-8526, Japan}
\email{adel.umirbayeva@gmail.com}

\author{Tetsuharu Maruta}
\affiliation{Physics Program, Graduate School of Advanced Science and Engineering, Hiroshima University, 1-3-1 Kagamiyama, Higashi-Hiroshima, Hiroshima 739-8526, Japan}
\email{adel.umirbayeva@gmail.com}

\author{Ryo Imazawa}
\affiliation{Physics Program, Graduate School of Advanced Science and Engineering, Hiroshima University, 1-3-1 Kagamiyama, Higashi-Hiroshima, Hiroshima 739-8526, Japan}
\email{adel.umirbayeva@gmail.com}

\author{Tatsuya Nakaoka}
\affiliation{Hiroshima Astrophysical Science Center, Hiroshima University, 1-3-1 Kagamiyama, Higashi-Hiroshima, Hiroshima 739-8526, Japan}
\email{adel.umirbayeva@gmail.com}

\author{Mahito Sasada}
\affiliation{Institute of Innovative Research, Tokyo Institute of Technology, 2-12-1 Ookayama, Meguro-ku, Tokyo 152-8551, Japan}
\email{adel.umirbayeva@gmail.com}

\begin{abstract}

We present a detailed study of the magnetic field structure in the {\cloudname} molecular cloud, a ring-like filamentary cloud within the NGC 7538 region. Our analysis combines multi-wavelength polarization data and molecular line observations to investigate the magnetic field's role in the cloud's formation and evolution. We utilized interstellar dust polarization from the {\planck} telescope to trace large-scale field orientations, starlight extinction polarization from the Kanata telescope to probe the cloud’s magnetic field after foreground subtraction, and velocity gradients derived from CO isotopologues observed with the IRAM 30m telescope to examine dense regions.
Our results reveal a coherent yet spatially varying magnetic field within {\cloudname}. The alignment between {\planck}-derived orientations and starlight extinction polarization highlights significant foreground dust contamination, which we correct through careful subtraction. The global alignment of the magnetic field with density structures suggests that the field is dynamically important in shaping the cloud.  Variations in CO-derived orientations further suggest that local dynamical effects, such as gravitational interactions and turbulence, influence the cloud’s structure. The curved magnetic field along the dense ridges, coinciding with mid-infrared emission in WISE data, indicates shock compression, likely driven by stellar feedback or supernova remnants.
Our findings support a scenario where {\cloudname}'s morphology results from turbulent shock-driven compression, rather than simple gravitational contraction. The interplay between magnetic fields and external forces is crucial in shaping molecular clouds and regulating star formation. Future high-resolution observations will be essential to further constrain the magnetic field's role in cloud evolution.

\end{abstract}


\section{Introduction} 
\label{sec:intro}

Star formation is a fundamental process within galaxies and filaments play a pivotal role in star formation, acting as the skeletal structure within molecular clouds where stars are born. 
The study of magnetic fields is closely intertwined with the exploration of filamentary structures \citep{planck2014-xix,planck2014-XXXIII,planck2016-XXXV,alina2019, pattle2023}.
Their interaction offers valuable information about the evolution of the interstellar medium (ISM) and the conditions necessary for star formation.

There exist several theoretical explanations of filament formation, which can be divided into the following major groups (see \cite{abe2021} for a more detailed comparison or \citet{pattle2023} for an alternative classification):

\begin{itemize}
    \item Self-gravitational fragmentation of sheet-like clouds \citep{nagai1998,balfour2015}), which can be triggered by cloud-cloud collision, expanding HII regions, \revision{supernovae explosions} or motions due to the galactic rotation.
    \item Turbulent motions. In works of \citet{stone1998}, \cite{padoan2001} or \cite{federrath2016}, turbulence was proposed to trigger shock and compression of sheet-like structures. According to \cite{hennebelle2013} or \cite{inoue2012}, turbulence can stretch the existing inhomogeneities of parental clouds. Here, one of the main debate points is the nature of turbulence.  
    \item Converging flows and cloud-cloud collision \citep{inoue2013}.
    \item Cloud contraction and self-gravitating accretion \citep{nakamura2008,heitsch2013}).
\end{itemize}

The proposed scenarios imply different configurations, where many factors, such as the density of the filaments compared to the environment, as well as magnetization, play role. For instance, in sub-alfv\'enic turbulence, where the magnetic field is a dominant factor, one would expect perpendicular alignment \revision{between the magnetic field and the local gas structures} \citep{barretomota2021} while in super-alfv\'enic turbulence, where the magnetic field is less dominant, different configurations can be possible \citep{mazzei2023}.
In addition, perpendicular relative orientation between the magnetic field and the filament can be expected in self-gravitational contraction \revision{of the filament itself}, where the contraction is most efficient along the magnetic field lines. 
In contrast, parallel relative orientation is expected for sub-alfv\'enic turbulence, and in the compressed sheets in the super-alfv\'enic regime \citep{hennebelle2013} or filaments formed by isothermal sheet fragmentation \citep{nagai1998}. Also, anisotropy of the magnetohydrodynamic (MHD) turbulence can lead to the formation of filaments parallel to the magnetic field \citep{xu2019} \revision{due to perpendicular turbulent mixing effects}.
Moreover, \revision{some filament formation mechanisms can affect not only the magnetic field versus density structures alignment but also the shape of the filaments themselves \citep{pattle2023}. For instance,} several filament formation mechanisms predict curved or ring-like structures of the filaments. Turbulent shear flows in molecular clouds can stretch dense structures \citep{hennebelle2013,inoue2016}, and the presence of a magnetic field can induce tension which would curve a filament. Also, shock waves from supernova explosions, stellar winds \citep{koo1992}, or cloud collisions can distort filaments, leading to curved shapes.   
Nonlinear fluid instabilities can create irregularities in the density. Those irregularities may evolve into curved structures \citep{lachieze1981,vishniac1994,stone2009}. 
\revision{}

\revision{Although the relative alignment can be alternated during the evolution of molecular clouds, we categorize the above-mentioned expected relative orientations between filament morphologies and magnetic fields into four groups:
\begin{enumerate}
    \item Filament parallel to the magnetic field would be typically expected in sub-Alfv\'enic turbulence or in shock-compressed layers.
    \item Filament perpendicular to the magnetic field would be typically expected in self-gravitational collapse along magnetic field lines or in super-Alfv\'enic turbulence.
    \item Curved magnetic field would be seen in ring-like or arc-shaped filaments potentially resulting from expansion by stellar feedback
    \item Complex relative orientation would be expected in highly turbulent or very dynamic environments.
\end{enumerate}
These scenarios are not exclusive and may co-exist within a single cloud}.
Theoretical studies of filament formation and alignment with magnetic fields reveal a variety of possible configurations, depending on factors such as the turbulence regime, magnetic field strength, gravitational influence, and formation history. However, the complexity and diversity of these scenarios underscore the need for observational data to validate and refine our understanding of these processes in real ISM.
Thus, individual studies of the morphology of molecular clouds and their magnetic field can provide valuable input 
and can be used to test theories. We propose to analyze the magnetic field structure of a ring-like filamentary cloud \cloudname. The cloud's striking geometry 
can be seen as a perfect test bench of the formation scenarios described above, and the information on the cloud's magnetic field structure may help to disentangle between them.



{\cloudname} is a part of a larger molecular cloud complex that includes active star-forming regions NGC 7538 and S159 \citep{brunt2003}. 
Previous studies of {\cloudname}'s star-forming activity have used spectroscopic and continuum observations. \cite{frieswijk2007} obtained C$^{18}$O (2-1) intensity maps which revealed a filamentary curved structure. 
They also studied the properties of the southern cores and the S159 region using NH$_3$, CS, and CO isotopologues observations, discovering intermediate and high-mass cores with temperatures indicative of  internal heating sources.
\cite{fallscheer2013} used Herschel observations to highlight the quasi-elliptical ring shape of {\cloudname} at wavelengths 250 $\mu$m and longer. Their spectral energy distribution (SED) fitting detected many compact sources, identifying eight high-mass dense clump candidates, mostly located in the southern part of {\cloudname}. These findings, together with previous molecular line studies \revisionn{\citep{frieswijk2007}}, suggest that {\cloudname} is a \revisionn{massive molecular cloud complex}, with an estimated total mass \revisionn{of the cores} of $\sim 3000\,M_\odot$, \revisionn{which} could potentially lead to cluster formation. \revisionn{In addition, the region's} structural complexity indicates that the cloud is relatively evolved.


However, the origin of the cloud formation and its specific shape, remains unknown. \cite{fenske2021} conducted a comprehensive study to investigate the origin of the elliptical ring structure. They explored various possible formation mechanisms, such as a stellar wind or a supernova explosion, and concluded that a star within NGC 7538 might have initiated the ring formation through an ejection event driven by an external force. 
Despite this, they did not find a promising candidate responsible for the formation of {\cloudname}. 
Given these uncertainties, another promising approach to study the formation and evolution history of {\cloudname} is to analyze its magnetic field structure. 
Magnetic fields can serve as tracers of the cloud's developmental processes, offering insights into the forces and dynamics that have shaped its current configuration \citep{li2014,alina2021}. 
By examining the magnetic field orientations within {\cloudname}, we can better understand the mechanisms that have influenced its evolution. 

Ring-like shape structures and their magnetic fields can be studied using sub-millimeter observations of the interstellar dust polarized emission from HII regions. For instance, \cite{fernandez-lopez2021} studied an ultracompact HII region G5.89-0.39 using the Atacama Large Millimeter and Submillimeter Array (ALMA) and found an irregular shape of the polarization pattern, from radial in the inner part to azimutal at the outer parts. However, the correspondence between polarization and magnetic fields is not straightforward because of the uncertainties on the alignment mechanisms' nature at the scale of the HII structure, which is around 10 000 au. Yet another example can be found in the study of \cite{konyves2021} which utilized James Clerk Maxwell Telescope (JCMT) polarization data to map the magnetic field structure of a region within the Rosette Molecular Cloud. The study revealed that the embedded parsec-size HII regions exhibited partly aligned magnetic fields. However, both studies investigated small, sub-pc and parsec scale structures. \revision{Similarly, \citet{tahani2023} analyzed the magnetic field morphology around compact HII regions and identified tangential field configurations relative to ring-like structures within sub-parsec scale regime.}
 
In contrast, our analysis focuses on a large, more than 10 parsec, elongated ring-like cloud with ongoing star formation activity indicating a stage beyond the pristine early star formation.
The magnetic field within the structure has not been previously investigated and can provide valuable insights on the past and ongoing physical processes.

Starlight polarization by interstellar dust grains and their polarized emission are commonly used to map the plane-of-the-sky (POS) interstellar magnetic field. In addition, the Velocity Gradients Technique (VGT) introduced by \cite{lazarian2018vgt} broadens the toolset of magnetic field studies. The technique utilizes spectroscopic data and has proved its efficiency in observational studies of molecular cloud regions and consistency with polarization data \citep{hu2018,hu2019,alina2021}. By combining these tools, we obtain a holistic probe of plane of the sky (POS) magnetic field structure in the ISM, where dust and gas are mixed.

\section{Data} 
\label{sec:style}

\subsection{Sub-millimeter Polarization data} 
\label{subsec:sum-mm-pol-data}

\begin{figure}
    \centering
    \includegraphics[width = 0.45\textwidth,trim = {2.5cm 0 2cm 0},clip]{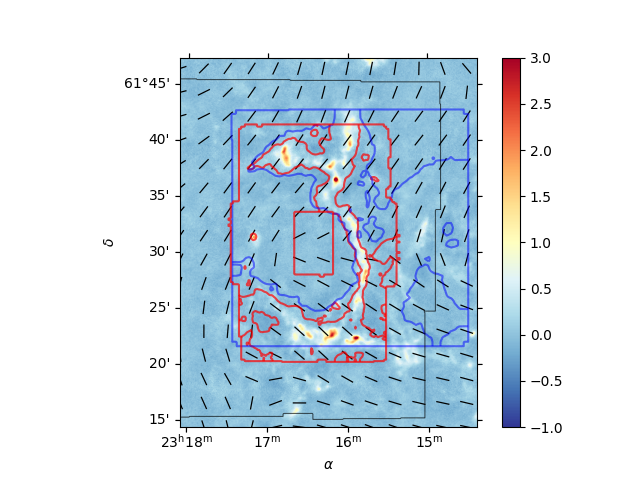}
    \caption{{\cloudname} molecular cloud. The background image \revision{shows} the JCMT 850 $\mu$m intensity map in mJy. \revision{The black segments show the magnetic field orientation in the POS, as inferred from {\planck} polarization data at the corresponding frequency at 353 GHz, rotated by 90$^{\circ}$} The blue and red rectangles are the footprints of the TRAO 14m and IRAM 30m observation fields, respectively. The red curve corresponds to the IRAM 30m {\coo} contours at 15 K km/s. The blue curve corresponds to the TRAO 14m {\co} contours at 95 K km/s. The black rectangle delimits the footprint of starlight polarization observations with the Kanata 1.5m telescope.}
    \label{fig:general}
\end{figure}

We employ the \textit{Planck}\footnote{\textit{Planck} (\url{http://www.esa.int/planck}) is an ESA science mission designed and completed by a collaboration of institutes directly funded by ESA Member States, NASA, and Canada.} satellite's PR3 data release at 353 GHz, the highest frequency polarization channel that would trace dust emission polarization. We smooth the data from the nominal $5 \arcmin$  resolution down to $7 \arcmin$  to improve the signal-to-noise ratio (SNR), taking into account the full noise covariance matrix according to the procedure described in \cite{planck2014-xix}. Note that we preserve the sampling of the nominal resolution and represent $1.7 \arcmin$-size pixels.
Figure~\ref{fig:general} shows POS magnetic field orientation derived from the {\planck} data by rotating polarization angles by $\pi /2$, as described further in Section~\ref{sec:polar}. 

\subsection{JCMT sub-millimeter emission data}
We use JCMT archival data at 850$\mu$m with the angular resolution of 14\arcsec. Observations covered the entire NGC7538 complex from which we extracted the {\cloudname} region. Figure~\ref{fig:general} shows the corresponding sub-millimeter dust emission map where we can distinguish the quasi-closed elliptical ring. The ring contains several dense compact structures which were previously detected by \cite{fallscheer2013} in \textit{Herschel} data. The map also features less dense and extended structures to the west from the ellipse's ridge, towards NGC7538, as well as to the southern part of the map.

\subsection{Starlight Polarization data}

\subsubsection{Observations at the Kanata telescope}
We obtained starlight polarization data in the H (1.6 $\mu$m) and R$_{\mathrm{C}}$ (0.65 $\mu$m, R band hereafter for simplicity) bands during observations at the 1.5m Kanata Telescope of the Higashi-Hiroshima Observatory in December 2023 and January 2024 using the Hiroshima Optical and Near-Infrared camera (HONIR; \citet{akitaya2014}). 
HONIR employs a cooled LiYF$_{4}$ Wollaston prism and is attached to the Cassegrain focus of the Kanata telescope.
Each observation set consisted of a sequence of exposures at four position angles (PAs) of the super-achromatic half-wave plate, $0\fdg 0$, $45\fdg 0$, $22\fdg 5$, and $67\fdg 5$. 
\revision{The exposure time of each frame is 75 seconds in the R band, and 60 seconds in the H band.}
A focal mask consisting of 5 rectangular apertures with each $0.75\arcmin \times 9.8\arcmin$ slots was used to prevent the overlap of the ordinary- and extraordinary-ray images.
\revision{A single field observation covers $7.8\arcmin \times 9.0\arcmin$ in the combined data.}
We ran each observation set with $3 \times 3$ spatial dithers, using a $31.2\arcsec$ step in the east-west direction and a $20.0\arcsec$ step in the north-south direction to cover each field by, at least, 3 observation sets.
\revision{Thus, most of the stars have at least three position data in 12 frames, giving actual exposure time of at least 900 seconds in the R band and 720 seconds in the H band.}
Due to the extend of {\cloudname}, we divided the region into nine overlapping fields with dimensions of $7.8\arcmin \times 9.0\arcmin$.
\revision{The angular resolution of the observations is 0.3$\arcsec$/pixel which is substantially less than the typical seeing of $2.5\arcmin$.}
The observation's coverage is represented in dark gray in Figure~\ref{fig:general}. 

The data were reduced by the HONIR's common pipeline.
The polarization parameters were calculated by the way described in \citet{kawabata1999}. 
We have confirmed that the instrumental polarization is negligible ($\Delta q, \Delta u \leq 0.1$\% in R band and $\leq 0.2$\% in H band) and the uncertainty of the PA of the polarization was less than 2$^{\circ}$ throughout the observation period by continuous calibration observation of the polarimetric standard stars \citep{schmidt1992}.
\revision{Typically, $\sigma_p$ reaches $0.2 \%$ at 15 mag in the R band.}
The pipeline provided the catalogs for each observation set containing pixel coordinates, polarization parameters $q=Q/I$ and $u=U/I$, and their observational errors.

\subsubsection{Inferring star distances using Gaia data}
We employed Gaia Early Data Release 3 catalog of star distances \citep{bailer-jones2021} from which we preliminary extracted the stars located within Kanata's FOV. In addition, we calculate the B, R, V, $\mathrm{I_c}$, J, H, and $\mathrm{K_s}$ bands magnitudes from the G band magnitudes and color information contained in Gaia data using Johnson-Cousins and 2MASS relationships\footnote{See \url{https://gea.esac.esa.int/archive/documentation/GDR3/Data_processing/chap_cu5pho/cu5pho_sec_photSystem/cu5pho_ssec_photRelations.html} for more details.}. Thus, our Gaia catalog contains information on distances and magnitudes in each band for each star in addition to the information contained in the original Gaia DR3.

We obtained the World Coordinate System (WCS) information for each observed image using astrometry.net platform. Based on this information, we determined the coordinates of each observed star and detected the corresponding stars in the Gaia catalog within a search radius of 
$1\arcsec$. By converting the observed flux to magnitudes for each observational band and comparing these with the magnitudes obtained from the Gaia catalog, we eliminated the possibility of mismatches.
Finally, this allows us to use Gaia's median photometric distance posterior estimate, known as \disgaia.

\subsubsection{Starlight polarization data selection}

\begin{figure}
    \centering
    \includegraphics[width=1.0\linewidth]{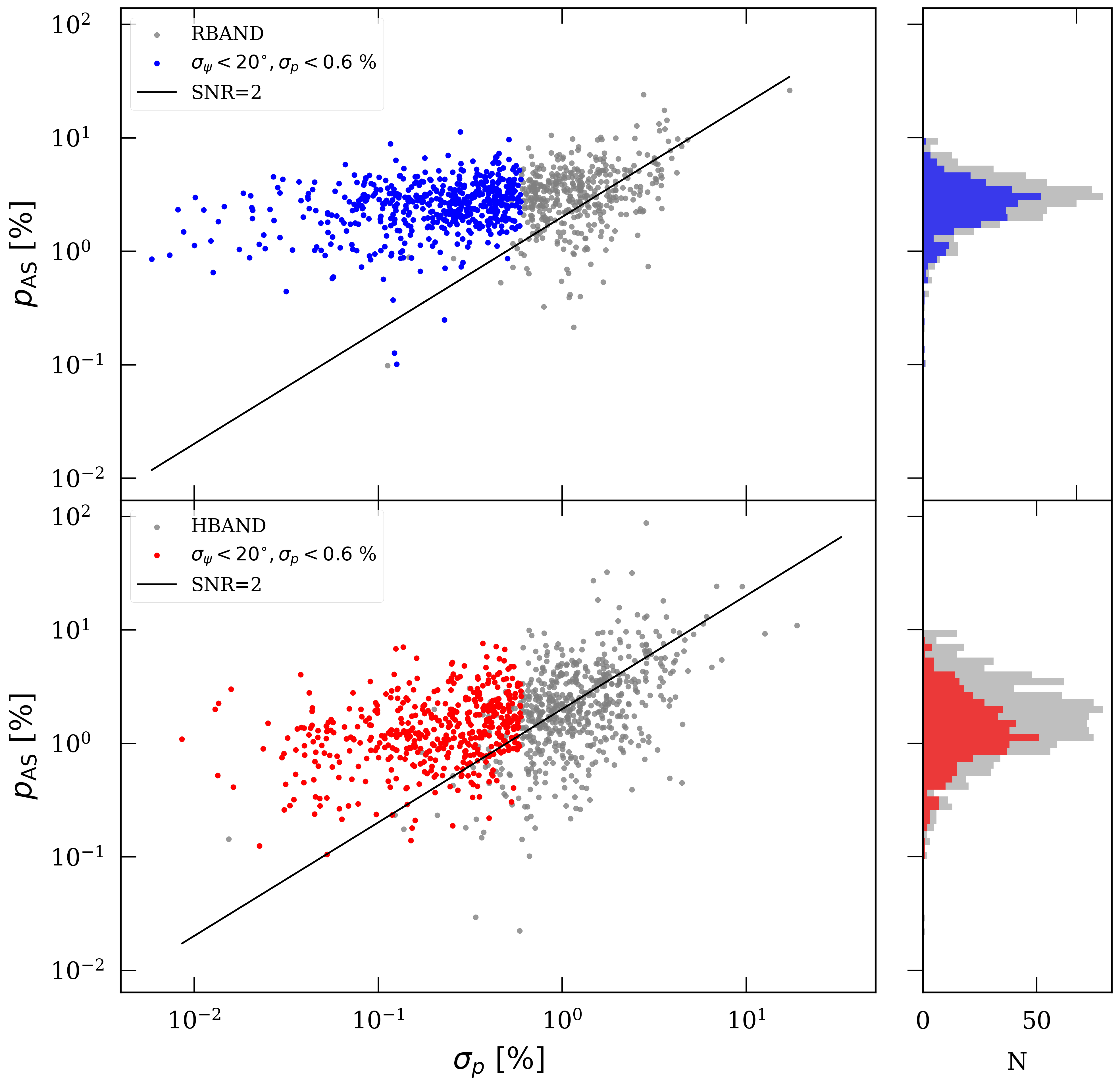}
    \caption{Left: Polarization fraction estimated using Eq.~\ref{eq:p_as} versus polarization fraction uncertainty ($\sigma_p$), \revision{based on the starlight polarization observations}, for the R band (top) and H band (bottom). The blue and red dots correspond to the data satisfying criteria based on the thresholds of $\sigma_p < 0.6\%$ and $\sigma_{Psi} < 20\%$. The red line shows SNR(p$_{AS}$) = 2. Right: distribution functions of $\sigma_p$.}
    \label{fig:snrp-new}
\end{figure}

Bias in polarization parameters (polarization fraction and angle) is a long-known issue \citep{serkowski1958,wardle1974,clarke1983} which can in principle be addressed if one knows the noise properties of the data \citep{Montier1,Montier2}. 
Starlight polarization measurements typically do not provide the full noise covariance matrix, a feature available at some sub-millimeter polarization instruments \citep{planck2014-xix,bernard2023}. Instead, it provides the variance of Stokes $Q$ and $U$ parameters. We employ the classical approach and compute uncertainties on polarization parameters as follows \citep{naghizadeh-khouei1993}:
\begin{equation}
    \sigma_p = \dfrac{\sqrt{Q^2 \sigma_Q^2 + U^2 \sigma_U^2}}{p}
\end{equation}
and
\begin{equation}
    \sigma_{\psi} = 28.65^{\circ} \dfrac{\sigma_p}{p} \,,
\end{equation}
where 
\begin{equation}
 p = \frac{\sqrt{Q^2 + U^2}}{I}
 \label{eq:p}
\end{equation}
 is the classical, or naive, estimator of polarization fraction.
In addition, we employ the asymptotic estimate of polarization fraction that accounts for the bias issue \revisionn{\citep{wardle1974}:}

\begin{equation}
    p_{AS} = \begin{cases} \sqrt{p^2 - \sigma_p^2} & p > 0 \\ 0 & p \leq \sigma_p \end{cases}
    \label{eq:p_as}
\end{equation}
\revision{This allows us to obtain the signal-to-noise ratio (SNR) value closer to the true SNR. In general, polarization studies use the SNR criterium to filter polarization data.} \cite{Montier2} \revision{showed that} the minimum threshold of at least at SNR = 2 on polarization fraction \revision{is necessary for most of the estimators while minimum SNR of three is necessary for the naive estimator} (Eq.\ref{eq:p}). However, limiting the SNR can lead to the elimination of low polarization fraction data and induce bias towards high polarization fraction values. 
\revision{Indeed, Figure~\ref{fig:snrp-new} shows that a cut based on SNR($p_{AS}$) would result in choosing measurements with high noise and high $p_{AS}$.}
Instead, we chose to base our selection on the uncertainties of polarization fraction and polarization angle, to directly address the quality of the measurements:
\begin{equation}
    \sigma_{\psi} < 20^{\circ} 
\end{equation}
and
\begin{equation}
    \sigma_{p} < 0.6 \% \, .
\end{equation}


Figure \ref{fig:snrp-new} shows that the $\sigma_p$ criterium largely corresponds to the threshold of SNR($p_{AS}$) larger than two for the R band. \revision{In both bands, our selection criteria are more restrictive than the the classic approach based on the SNR of polarization fraction.}
In addition, as can be observed in Figure~\ref{fig:snrp-new}, it also allows us to remove high, larger than 10$\%$, polarization fraction data, which are very likely to be caused by bias. Also, $\sigma_{\psi}$ criterium balances the lower wing of the distribution of $p$, which can be due to the averaging of uncertain measurements.  
As expected, the H band shows higher noise due to the fainter signal. 

\subsection{Molecular Line data}
\subsubsection{TRAO 14m telescope data}
{\cloudname} has been observed as part of the "TOP" key science program at the Taeduk Radio Astronomy Observatory (TRAO)\footnote{\url{https://trao.kasi.re.kr/main.php}}. This survey observed 2000 Planck Galactic Cold Cores using CO (J=1-0) transitions \citep{liu2018b}. 
The angular resolution of the data is 47\arcsec, and the spectral resolution after smoothing is 0.3 km/s.
The field observed by the TRAO 14m telescope towards {\cloudname} is shown in blue in Figure~\ref{fig:general}, and the contours correspond to 95 K km/s integrated intensity level.  

\subsubsection{Observations at the IRAM 30m telescope}
We observed {\cloudname} at the Institut de Radio Astronomie Millim\'etrique (IRAM) 30m telescope in the period between January and March 2022 (project ID: 125-22). The Eight MIxer Receiver instrument (EMIR\footnote{\url{https://publicwiki.iram.es/EmirforAstronomers.}}) at the 3 mm band was tuned at 109.7 GHz in combination with the fast Fourier transform spectrometers (FTS) at 50 KHz resolutions that yielded $\approx 0.5$ km s$^{-1}$ resolution at the chosen band. This allowed us to obtain $^{13}$CO and C$^{18}$O emission lines. We mapped the entire region by separating it into 6 slightly overlapping tiles covering the elliptical ring using the on-the-fly (OTF) mapping mode with an OFF position outside the molecular complex. The footprint of the observed field as well as the contours of the {\coo}-integrated intensity at 15 K km/s, are shown in red in Figure~\ref{fig:general}.
We employed the Continuum and Line Analysis Single-dish Software (CLASS\footnote{\url{https://www.iram.fr/IRAMFR/GILDAS/doc/html/class-html/class.html}}) for the data reduction and generation of spectral cube files. 

\subsection{Deriving magnetic field angle using interstellar dust polarization}
\label{sec:polar}
Elongated interstellar dust grains tend to gain rotation around their shortest axis and align it with the local magnetic field's direction \citep{vaillancourt2007}. However, the exact origin for alignment is not fully understood and may vary depending on the physical conditions \citep{dolginov1972,purcell1979,draine2013,lazarian2007,hoang2018}. The linear dichroism of the interstellar dust grains, or difference in absorption properties along different axes, is fundamental to understand the origin of the linear polarization. Unpolarized starlight is preferentially absorbed along the longest axis of the grains, while light polarized perpendicular to this axis can pass through. In the sub-millimeter range, thermal emission from the grains is stronger on their largest surface and is polarized parallel to the longest axis. Thus, dust polarization observations allow us to trace the POS interstellar magnetic field direction: directly in the visible and by rotating by 90$^{\circ}$ in the sub-millimeter domain.

Starlight polarization by dust extinction in the visible using the Kanata telescope data is calculated through the following equation:
\begin{equation}
    \psi_V = 0.5 \arctan(U,Q) \, .
    \label{eq:vis}
\end{equation}
Here, $0$ is toward the North, and increases counterclockwise, according to the International Astronomical Union (IAU) convention. In what follows, we will denote the POS magnetic field angles derived using the R and H bands data as {\bvis} and {\bnir}, respectively.

The {\planck} dust emission polarization is calculated using:
\begin{equation}
    \psi_{S} = 0.5 \arctan(-U,Q) \, .
        \label{eq:submm}
\end{equation}
The negative value of the Stokes $U$ parameter facilitates the conversion from the COSMO convention of the {\planck} data to the IAU convention. To obtain the magnetic field angle we rotate $\psi_{S}$ by 90$^{\circ}$.
In the following, we denote the corresponding POS magnetic field angle as {\bplanck}.


\subsection{Subtracting foreground polarization}
\label{sec:fgsubtraction}
The interstellar dust polarized emission observations in the sub-millimeter range, as well as the starlight polarization by extinction in the visible, correspond to an optically thin emission regime, which results in a signal integrated over the line of sight. 
Therefore, to isolate polarization originating from a distant cloud, it is essential to subtract both the foreground and background contributions. In the case of the dust polarization in the visible domain, where the distances to the observed stars are known, only the foreground contribution needs to be removed. 

Since star observations are "pencil-like" meaning that they do not form a regular gridded distribution and do not have a spatial resolution, we perform interpolation between data points across the map. We apply the nearest-neighbor interpolation method to process the polarization data of the molecular cloud. This method involves assigning each unknown value on the target grid the value of the nearest known data point and minimizes the distance between the target and known points. This allows us to fill in gaps in the data without creating artificial artifacts that may arise from using more complex interpolation methods. 
\revision{This approach was chosen specifically because the spatial distribution of stars in our data is highly non-uniform: some regions have dense sampling, while others contain large gaps. Using linear or spline-based interpolation in such sparse regions would introduce artificial structures or smooth values into physically unconstrained zones, which we aimed to avoid. }
The spatial interpolation is performed for the Stokes Q and U parameters. We first define the distance that constitutes the foreground, $\mathrm{d}_{\mathrm{fg}}$. Second, by interpolating data of the stars located within the range [0, $\mathrm{d}_{\mathrm{fg}}$] pc, we obtain $Q_{\mathrm {fg}}$ and $U_{\mathrm {fg}}$. Thus, a regularly gridded foreground map allows us subtract the foreground contribution and obtain the cloud polarization as follows:
\begin{eqnarray}
    Q_{\rm{cloud}} &=& Q_{\rm{d>\mathrm{d}_{\mathrm{cloud}}}} - Q_{\rm{d<\mathrm{d}_{\mathrm{fg}}}} \, , \\
    U_{\rm{cloud}} &=& U_{\rm{d>\mathrm{d}_{\mathrm{cloud}}}} - U_{\rm{d<\mathrm{d}_{\mathrm{fg}}}} \, .
\end{eqnarray}
The result is then converted to polarization angle using Eq.~\ref{eq:vis}.
\revision{Note that an alternative method of smoothing starlight polarization to the Planck resolution has been tested, and described in Appendix~\ref{sec:app}. The interpolation technique described above is more advantageous as it provides us with a uniform grid that is used in foreground contribution subtraction, although smoothing only removes several pixels from the map, as can be seen in Figure~\ref{fig:appendix_am_planck-starpol}.}

\subsection{Deriving magnetic field angle using Velocity Gradients Technique}

Based on the theory of magnetohydrodynamic turbulence by \citet{goldreich1995,LV99}, \citealt{GL17,LY18a,HYL18} developed the velocity gradient technique (VGT) to trace the magnetic field structure using spectroscopic data. 
\revision{The technique relies on the anisotropy of MHD turbulence, in which turbulent eddies are elongated along magnetic field lines. This anisotropic structure implies that the velocity gradient tends to be oriented perpendicular to the magnetic field, making it a useful tracer of magnetic field orientation in turbulence-dominated regions. This implies computing the velocity gradients rotated by 90$^{\circ}$.
However, when gravitational collapse or other non-turbulent dynamical processes become significant, the relative orientation between the velocity gradient and the magnetic field may change. Thus, the technique can be used to molecular clouds supported by turbulence.}
The method can be applied to two-dimensional intensity or velocity maps. 
\revision{To extract velocity information from spectroscopic data, one can use either velocity centroid maps C($x, y$) or velocity channel maps Ch($x, y$).  They are calculated from:}
\begin{eqnarray}
C(x, y) &=& \frac{\int \rho(x, y, v_{\rm los}) v_{\rm los} dv_{\rm los}}{\int \rho(x, y, v_{\rm los}) dv_{\rm los}}, \\
 Ch(x,y) &=& \int^{v_0+\Delta v/2}_{v_0-\Delta v/2} \rho(x,y,v_{\rm los})dv_{\rm los},
\end{eqnarray}
where $\rho(x,y,v_{\rm los})$ represents the intensity values within the PPV cube, and 
$v_{\rm los}$ is the LOS velocity.
$v_0$ is the center velocity of the channel and $\Delta v$ is the width of the channel, chosen to be less than the square root of the dispersion of the velocity of turbulence, $\Delta v<\sqrt{(\delta v)^2}$.
\revision{These two maps have different weights of density and velocity information. Here, we use the velocity channel gradients (VChGs), since it has more velocity information due to the velocity caustics effect \citep{LP00} and allows us to trace the magnetic field in specific velocity ranges.
In addition,} the concept of the connection between local magnetic fields and gradients is only valid statistically, and we use sub-block averaging, as suggested by \citet{yuen2017vgt}, with the blocks of 20$\times$20 pixels. 
\revision{This results in obtaining the final angular resolution of $5\arcmin$, comparable to the resolution of the {\planck} data}.
We only consider pixels with at least 95$\%$ of the sub-block population exceeding the mean rms noise. 
We applied the velocity gradients technique to the CO-isotopologues data ({\co}, {\coo}, {\cooo}), and denote the derived magnetic field as {\bvgt}.

\subsection{Alignment Measure}
The different methods used in this study to derive the magnetic field structure have different capabilities and limitations due to the different physical phenomena involved. 
We see those as an advantage and make use of their synergy.
Further, we employ the alignment measure parameter (AM) to compare the magnetic field geometries:
\begin{equation}
    \mathrm{AM} = 2 \, (\cos^2{ \theta - \frac{1}{2}} )\, , 
\end{equation}
where $\theta = |B_1 - B_2|$, and $1$ or $2$ subscript denote different methods.
AM = $-1$ indicates a perpendicular relative orientation, while AM = $1$ indicates a parallel relative orientation.
\section{Results}


\subsection{Magnetic field structure derived from {\planck} measurements}
Despite the low angular resolution of the {\planck} data, \revision{the magnetic field map in} Figure~\ref{fig:general} shows the variation of the magnetic field orientations over the observed region. In particular, three main uniform directions can be distinguished: 
\begin{itemize}
\item a quasi-horizontal direction in the southwestern and central part, which coincides with the galactic plane orientation;
\item a quasi-vertical direction in the southeastern part of the map;
\item a SE to NW orientation in the northern part.
\end{itemize}
Those changes in the magnetic field orientation confirm that the {\cloudname} filament, and the NGC7538 region in general, have a distinct magnetic field structure which stands out from the Galactic plane magnetic field, and its dust polarized emission has a significant contribution to the sub-millimeter signal. 
However, the sub-millimeter observations alone, at the {\planck} angular resolution, do not allow us to access the magnetic field structures of the different components along the line of sight, and we employ our starlight extinction observations.

\subsection{Magnetic field structure derived from optical polarization measurements}

\subsubsection{Foreground polarization angle}

Although the distance to {\cloudname} has not been specifically measured, it is associated with the NGC7538 star-forming region by its velocity properties. The latter was estimated between 2700$\pm$100 pc by \cite{moscadelli2009} using trigonometric parallaxes and 3100$\pm$1200 pc \citep{brand1993} using spectrophotometric distances of HII regions and fitting to the galactic rotation curve. Here we adopt the most recent distance of 2700 pc, which was also used in \cite{fallscheer2013}.
Taking into consideration this estimate and the distribution of the photometric magnitudes described above, we set the upper limit for the foreground to $\mathrm{d}_{\mathrm{fg}} = ${\fgd} pc. 

\begin{figure}
    \centering
    \includegraphics[width=1\linewidth]{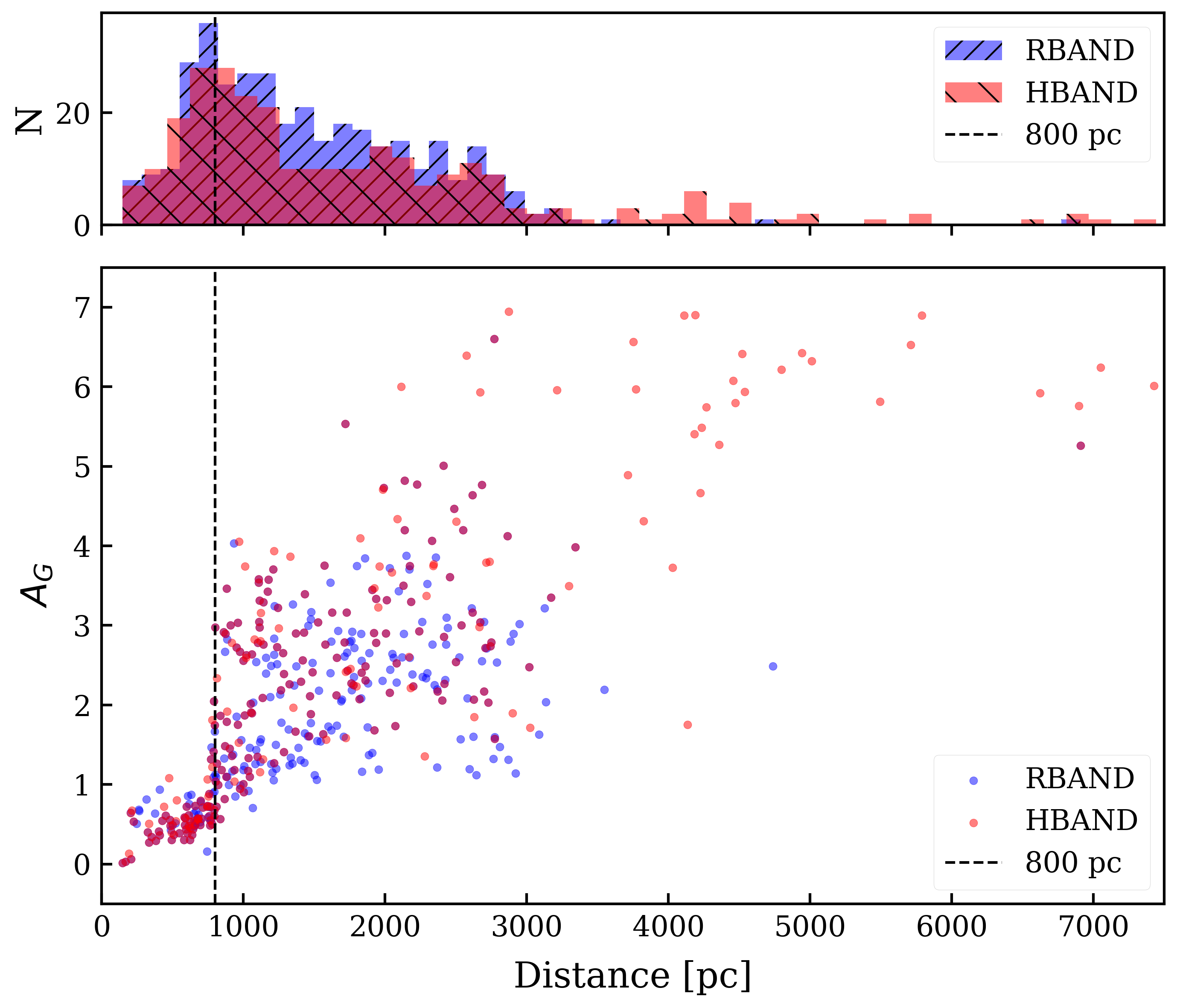}
    \caption{Distribution of the number of stars (N) and their extinction (A$_G$) as a function of distance in pc ({\disgaia} in the Gaia catalog) in the R and H bands. The vertical dashed line indicates a significant increase in the number of stars at $\simeq 800$ pc.} 

    \label{fig:avh-avr}
\end{figure}

\begin{figure}
    \centering
    \includegraphics[width=1\linewidth]{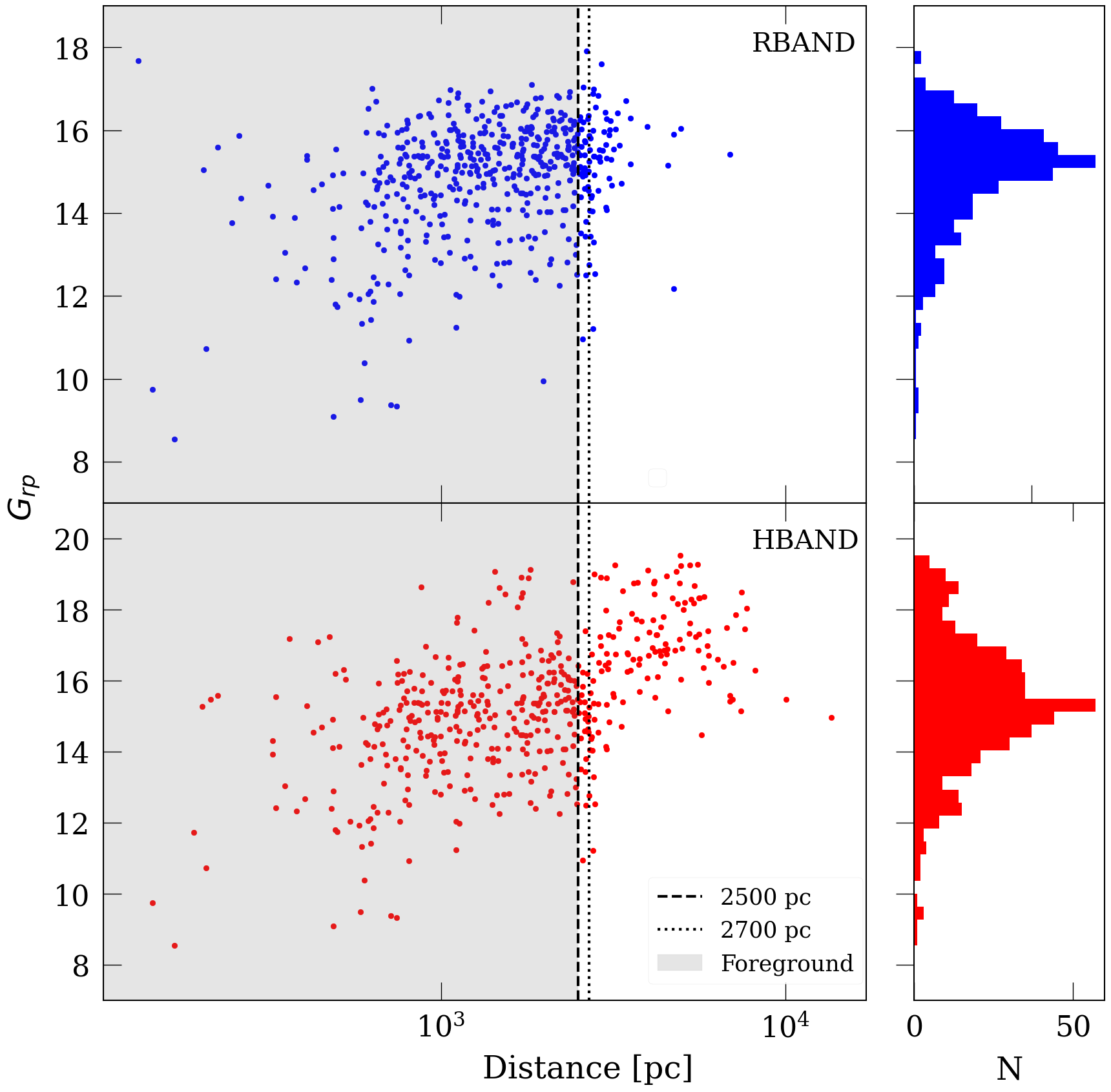}
    \caption {Distribution of the photometric magnitude in the Gaia RP band (G$_{rp}$) as a function of distance in pc for the stars observed in the R (upper panel) and H (lower panel) polarimetric bands. 
    The vertical dashed and dotted lines correspond to distances of $2500$ pc and $2700$ pc, respectively. The red-colored area shows the stars that were used for the foreground estimation and subtraction.}
    \label{fig:avh-avr-sc} 
\end{figure}

Figure~\ref{fig:avh-avr} shows the interstellar extinction of the stars observed in both HONIR bands toward {\cloudname}. 
A step-like rise of the upper boundary of the extinction in the G band at $\simeq 800$ pc indicates about non-negligible foreground interstellar material in the direction of the cloud. In addition, the lower boundary of the interstellar extinction increases more rapidly after that distance.
We also observe a less prominent increase of $\mathrm{A_{G}}$ at around 1700 pc.
Figure~\ref{fig:avh-avr-sc} shows the Gaia magnitude in the red photometric band $\mathrm{G_{rp}}$ of these stars as a function of distance, in R and H bands. Here, we also notice a significant increase of around 2 mag at $\simeq 800$ pc, which is more prominent in the H band. 
In addition, the density of the data points increases beyond the distance of $\simeq 2500$ pc, and the lower boundary of the magnitudes increases abruptly after $2700$ pc. 
Thus, notable changes in the dust distribution or in star population are observed at the distances of 800 pc and 2700 pc.

The right column of Figure~\ref{fig:avh-avr-sc} shows the probability distribution function of the magnitudes in both bands. Naturally, as the brightness of the stars decreases with the wavelength, the values are higher in the H band, with maximum at 19.5 mag. In the R band, the final dataset do not contain stars fainter than 18 mag. 
Initially, the dataset comprised 986 polarization measurements in the R band and 1265 in the H band. After applying all the thresholds, the final dataset consists of 528 measurements in the R band and 488 measurements in the H band.

\begin{figure}
    \centering
    \includegraphics[width=1\linewidth]{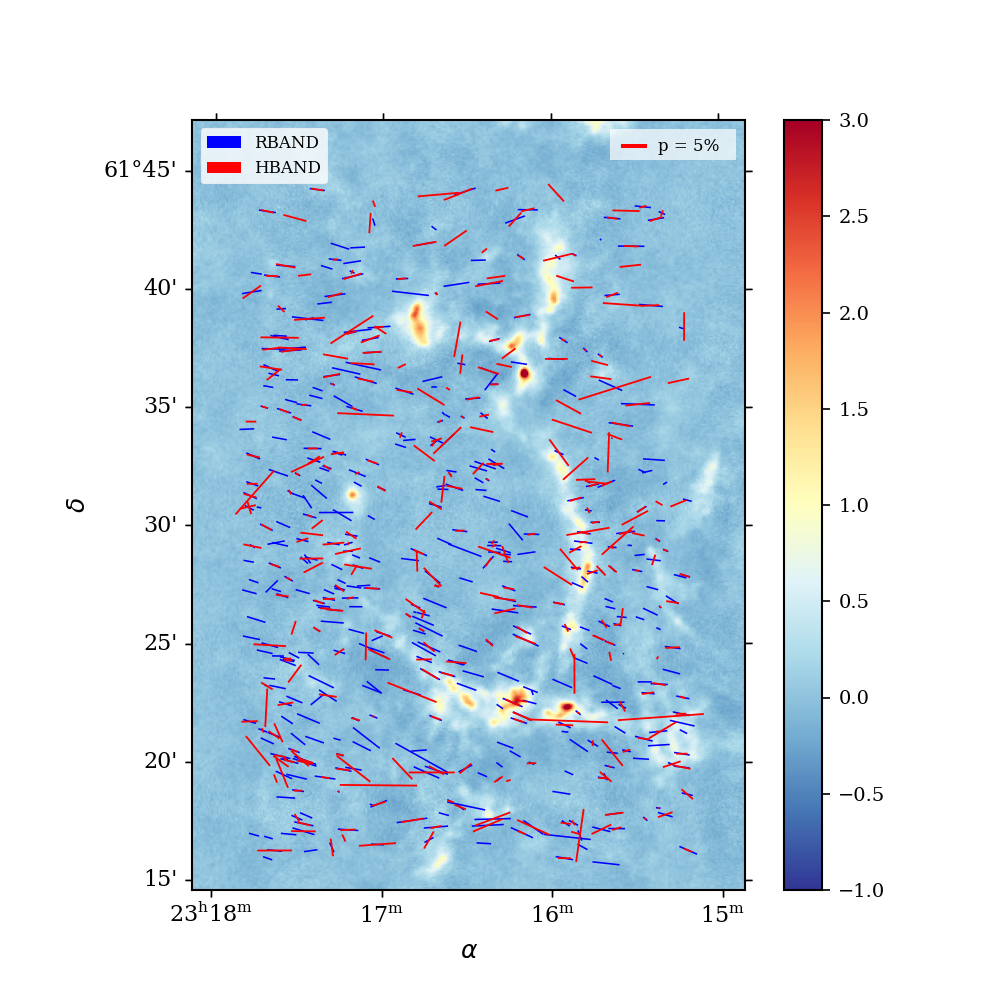}
    \caption{Starlight polarization in the H band (red bi-vectors) and R band (blue bi-vectors) for foreground stars with distances less than 2500 pc. The background image is the JCMT 850 $\mu$m intensity map in mJy. The reference of the bi-vector length for the polarization fraction of 5$\%$ is shown in the upper right corner.}
    \label{fig:fgstars}
\end{figure}

\begin{figure}
    \centering
    \includegraphics[width=1\linewidth]{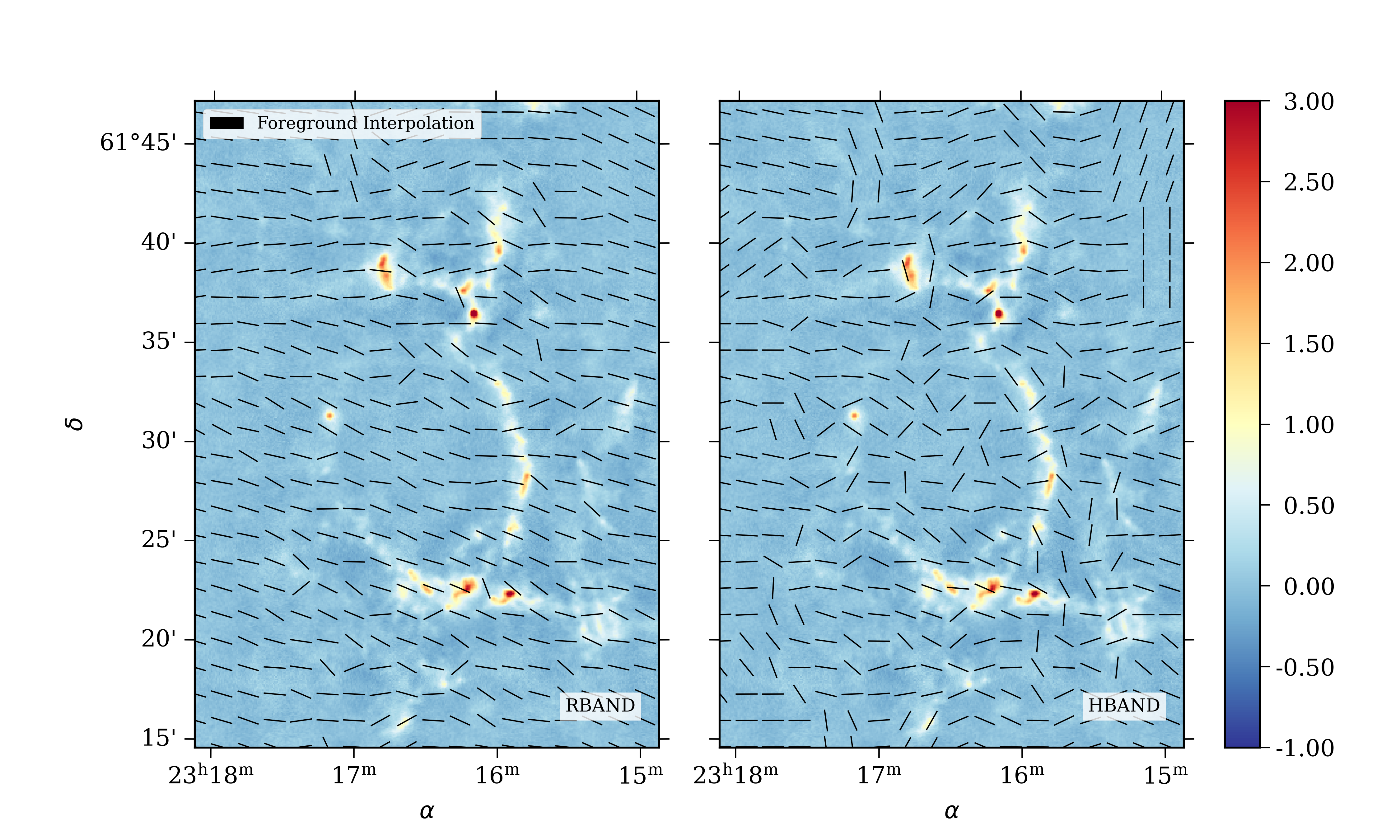}  
    \caption{Interpolated foreground magnetic field obtained from starlight polarization in the H (right) and R (left) bands toward {\cloudname}, using polarization data shown in Figure~\ref{fig:fgstars}. The background image represents the JCMT 850 $\mu$m intensity map in mJy.}
    \label{fig:fginterp}
\end{figure}

To represent the foreground magnetic field, we show in Figure~\ref{fig:fgstars} the starlight polarization observations for stars with distances less than 2500 pc and satisfying the criteria described above. Overall, we observe similar orientations of the bi-vectors. Their lengths correspond to the polarization fractions, and variations in the H band are higher, probably reflecting the more significant noise in the fainter band, and indicative of the sensitivity to interstellar structure variations along the line of sight.

Figure~\ref{fig:fginterp} represents the corresponding interpolated foreground magnetic field structure, obtained according to the procedure described in Section~\ref{sec:fgsubtraction}. 
The R band shows a rather uniform E-W orientation with a slight tendency for NE-SW diagonal. 
The H band polarization map with its similar E-W orientation exhibits more small-scale variations.
Both foreground {\bvis} and {\bnir} seem to rotate and align with the location of faint dust structures south and northeast of the ring. Interestingly, the curved filamentary structure south of the ring appears to be tightly aligned with the magnetic field. The presence of a twist in both wavelengths indicates that these faint structures are part of the foreground material \revision{as consistent polarization patterns imply similar contribution of dust to starlight polarization. If the structures were part of a more distant or spatially extended background component, such coherence would be less likely given the differential extinction and alignment effects. However, we cannot entirely exclude the possibility of a large-scale structure with coherent magnetic geometry projected along the line of sight.} 

\begin{figure}
    \centering
    \includegraphics[width=1\linewidth]{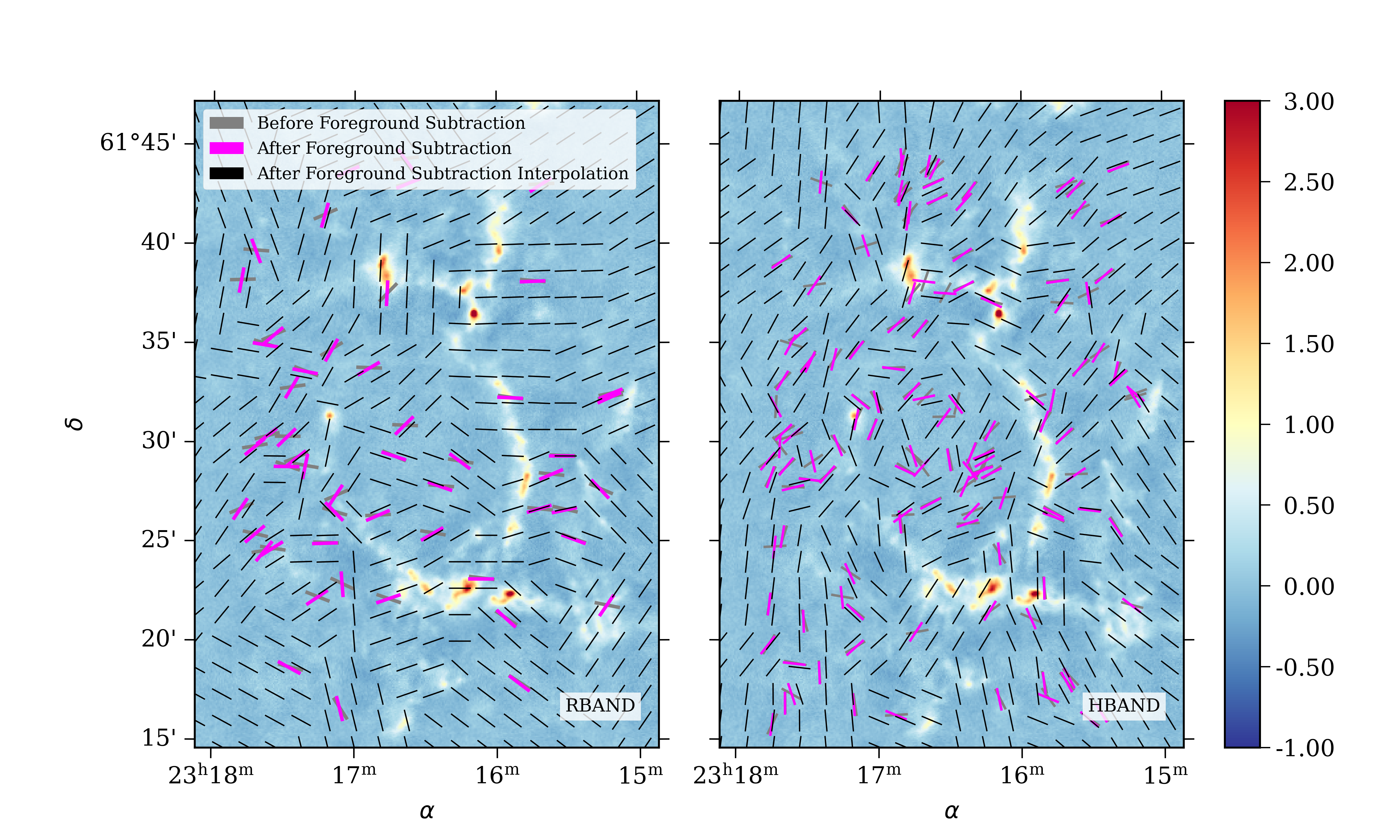}    
    \caption{Magnetic field structure obtained from starlight polarization in the H (right) and R (left) bands in {\cloudname}, after foreground subtraction (magenta segments), and the corresponding interpolated structure (black segments) The background image represents the JCMT 850 $\mu$m intensity map in mJy.}
    \label{fig:fgsubtraction}
\end{figure}

\subsubsection{Cloud polarization angle}

We subtract the foreground polarization component using the procedure described in Section~\ref{sec:fgsubtraction}. Figure \ref{fig:fgsubtraction} shows the resulting interpolated {\bvis} and {\bnir}. In the R band, the deviations from the global E-W orientations coincide with the higher intensity locations, such as the cores in the northern horizontal filament and in the southern far end of the eastern vertical filament. The H band has more fluctuations. However, we also observe the rotation of {\bnir} in the vicinity of the same regions.

The sparsity of the data of starlight polarization by extinction, limits the ability to trace variations of the magnetic field in the cloud. In addition, the near-infrared and the visible bands probe low-to-intermediate interstellar extinction material. We complement our study by using the velocity gradients technique.

\subsection{Magnetic field structure derived using the VGT on {\co} data}
\label{sec:trao}

\begin{figure}
    \includegraphics[width = \linewidth]{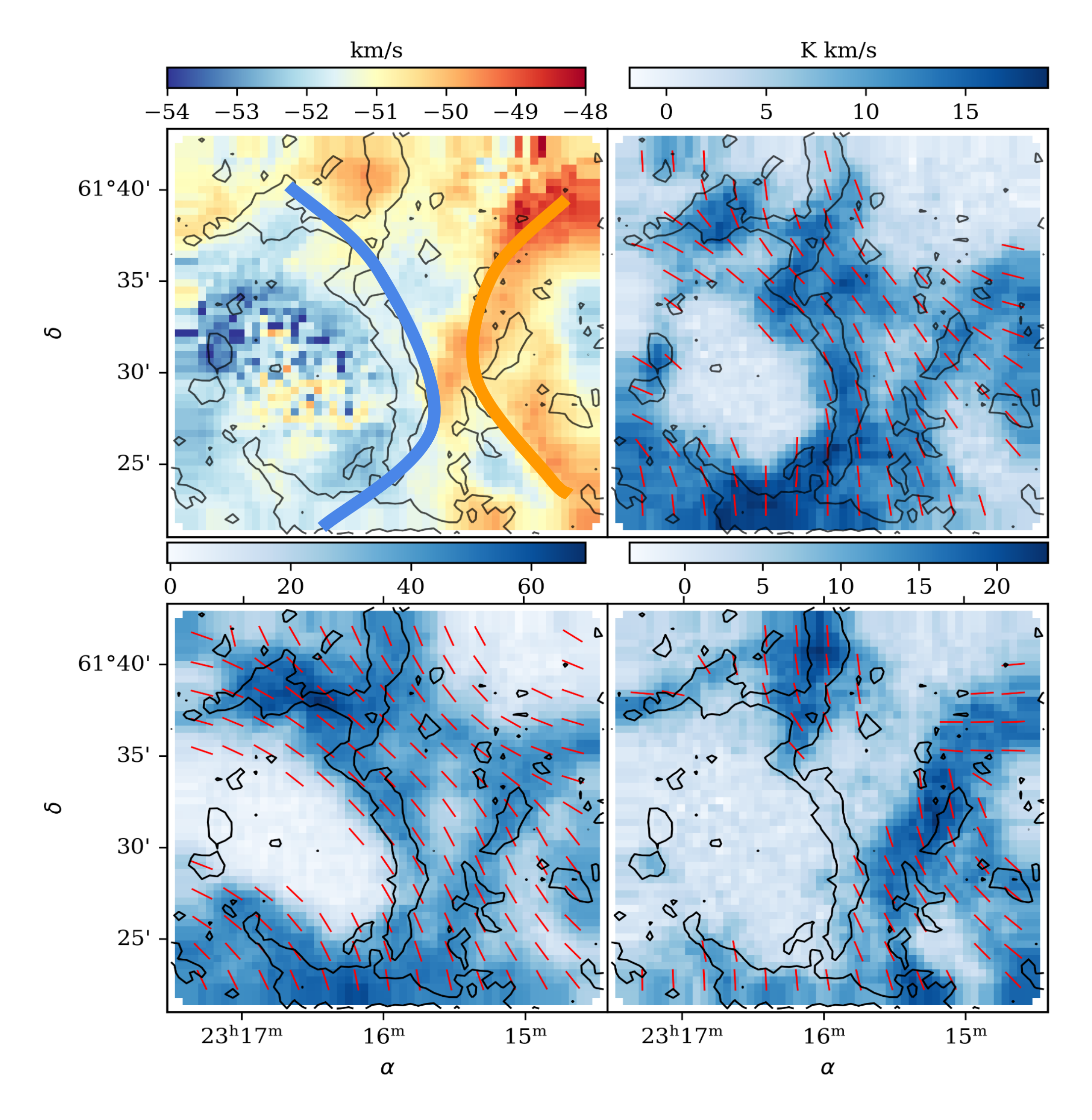}
        \caption{Maps of {\co} toward {\cloudname} from the TRAO 14m telescope observations. Top left panel: Velocity centroid (moment 1) map, in km/s, computed over the velocity range [-59;~-42] km/s. \revision{The blue and orange curves depict the two arcs described in Sect.~\ref{sec:trao}}. Top right: integrated intensity (moment 0) map in K km/s computed over the velocity range [-53.6;~-53] km/s. Bottom left: same as top right, over the velocity range [-52.5;~-50] km/s. Bottom right: same as top right, over the velocity range [-49.3;~-48.4] km/s. Red bi-vectors show the orientation of the local magnetic field derived using the velocity gradients technique, i.e. the orientation of the velocity gradients rotated by 90$^{\circ}$. Black contours correspond to the JCMT 850 $\mu$m  intensity contours 0.1 mJy.}
    \label{fig:trao}
\end{figure}

The TRAO $^{12}$CO (J=(1-0)) data allows us to trace the envelope and the outer regions of the cloud. The blue contours in Figure \ref{fig:general}, which correspond to {\co} contours, are evocative of two arcs located side by side, touching each other with the convex parts. The top left panel of Figure \ref{fig:trao} shows the {\co} velocity map. Indeed, two curved velocity coherent structures can be distinguished in the eastern and western parts, with an overall velocity gradient from SE to NW of the observed region. 
The rest of the panels of Figure~\ref{fig:trao} depict the orientation of the local magnetic field derived using the VGT over three consecutive velocity intervals: [-53.6;~-53] km/s; [-52.5;~-50] km/s, and [-48.4;~-46.5] km/s, for top right, bottom left and bottom right, respectively. Those ranges were chosen to show the maximum emission of the two curved structures and their common velocity interval. The western ring, in the bottom right panel, exhibits a rather intriguing {\bvgt} structure, which seems to follow the arc-like structure traced by the gas emission.  In this analysis, we mainly focus on the eastern ring. There, we observe a smooth rotation of {\bvgt} along the filament. However, the sensitivity of TRAO observations and its angular resolution limit further investigations.

\subsection{Magnetic field structure derived using the VGT on $\mathrm{^{13}CO}$ and $\mathrm{C^{18}O}$ data}
\label{sec:iram}
\begin{figure*}
    \centering
    \includegraphics[width=\textwidth]{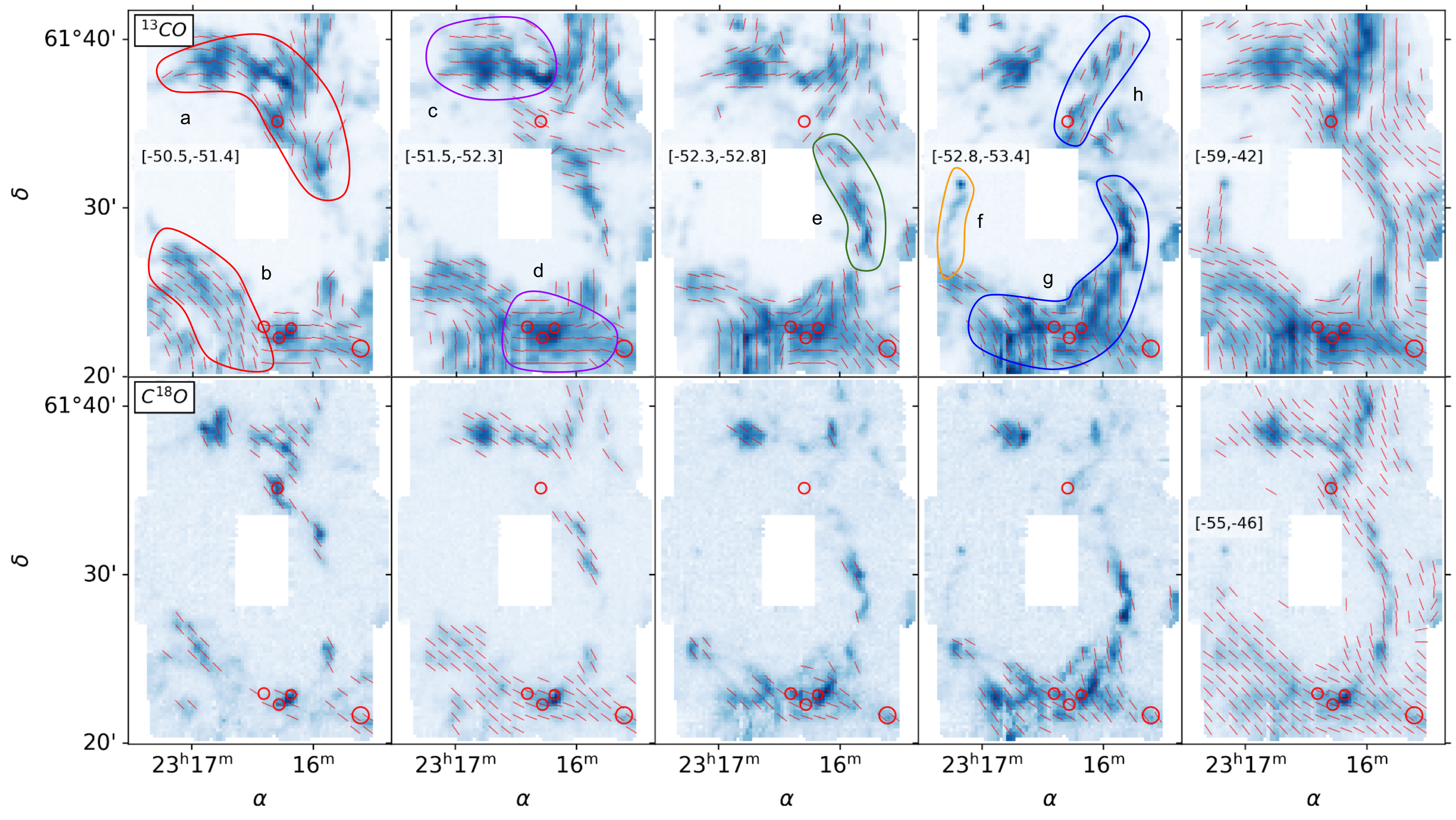}
    \caption{Magnetic field orientations (red segments) derived using the velocity gradients technique applied to {\coo} (upper row) and {\cooo} (lower row) IRAM 30m data. {\bvgt} are derived for the different velocity ranges: {\vrangemin}, {\vrangemid}, {\vrangemax}, and {\vrangelast} km/s from left to right, respectively. 
    In addition, {\bvgt} obtained from the entire velocity range in which {\cloudname} is detected, is represented in the last column. The red circles indicate the location of the high-mass dense clumps candidate objects reported by \cite{fallscheer2013} (see their Table 1). The circles' radii show their relative sizes with the smallest and the biggest corresponding to 0.4 and 1.1 pc. 
    The backgrounds are the integrated intensity maps over the indicated velocity ranges. 
    We represent {\bvgt} for pixels with lower thresholds of 2 K km/s and 0.5 K km/s for {\coo} and {\cooo}, respectively, except for the upper right panel where the threshold is taken at 2 K km/s.
    The letters \textit{a}-\textit{h} and the colored contours indicate the structures listed in Sect.~\ref{sec:iram}.}
    \label{fig:iram}
\end{figure*}

The IRAM 30m telescope observations ({\coo}(J=1-0) and {\cooo}(J=1-0)) allow us to trace denser parts of the clouds with higher angular resolution.
Structures that we associate with the {\cloudname} cloud span over the wide velocity range from -59 km/s to -42 km/s and exhibit a clear filamentary structure that follows the dust filaments observed with the JCMT as can be seen from the red contours in Figure~\ref{fig:general}. We distinguish several intensity components at different velocities both in {\coo} and {\cooo} maps in Figure~\ref{fig:iram}, at the same velocity intervals. The most prominent components, which are clearly seen both in dust and gas emission maps, are:

\begin{itemize}
    \item Two curved diagonal structures, opposite to each other, in the northern \revision{(denoted by \textit{a} in} Figure~\ref{fig:iram}) and southern \revision{(\textit{b})} parts at [{\vmin}; {\vmax}] km/s. The cloud appears more elliptical in this velocity range;
    \item Horizontal structures in the northern \revision{(\textit{c})} and southern \revision{(\textit{d})} parts at [{\vminn}; {\vmaxx}] km/s with the highest intensity values in the clumps;
    \item A thin vertical filamentary ridge in the western part \revision{(\textit{e})}, partially present at all velocity ranges; 
    \item A fainter and smaller vertical filamentary ridge \revision{(\textit{f})} in the eastern part at velocities under -53.4 km/s;
    \item Southern half-ring \revision{(\textit{g})} and a diagonal bar \revision{(\textit{h})} in the northwest at [{\vminnn}; {\vmaxxx}] km/s with the northern clump still present at these velocities.
\end{itemize}

In general, we observe a tight alignment of {\bvgt} with the above-described filamentary structures for given velocity intervals, for both tracers. 
The last column of Figure~\ref{fig:iram} shows that the overall {\bvgt} globally follows the elliptical ring geometry regardless of the molecular gas tracer.
A coherent curved geometry is observed in the western vertical ridge \revision{\textit{e}} across different velocity ranges.
In the northern part \revision{(\textit{a, c, h})}, the magnetic field orientation changes with the intensity structure, and, interestingly, a high-mass dense clump candidate has been identified previously by \cite{fallscheer2013} at the intersection of the two structures. 
In the northern clump \revision{\textit{c}}, there are clear changes in the {\coo}-{\bvgt} between the red-shifted and the blue-shifted parts. In addition, the different tracers show perpendicular orientations in the northern clump at {\vrangelast} km/s velocity range. The possible origin of such a behavior will be discussed further in Section~\ref{sec:disc_vgt}.
In the southern part of the cloud, the location of three candidate high-mass dense clumps coincides with the different structures intersecting each other with the respective aligned magnetic fields \revision{(\textit{b, d, g})}. 



\subsection{The warm dust structure and the magnetic field structure}

\begin{figure*}
    \centering
    \includegraphics[width=\textwidth]{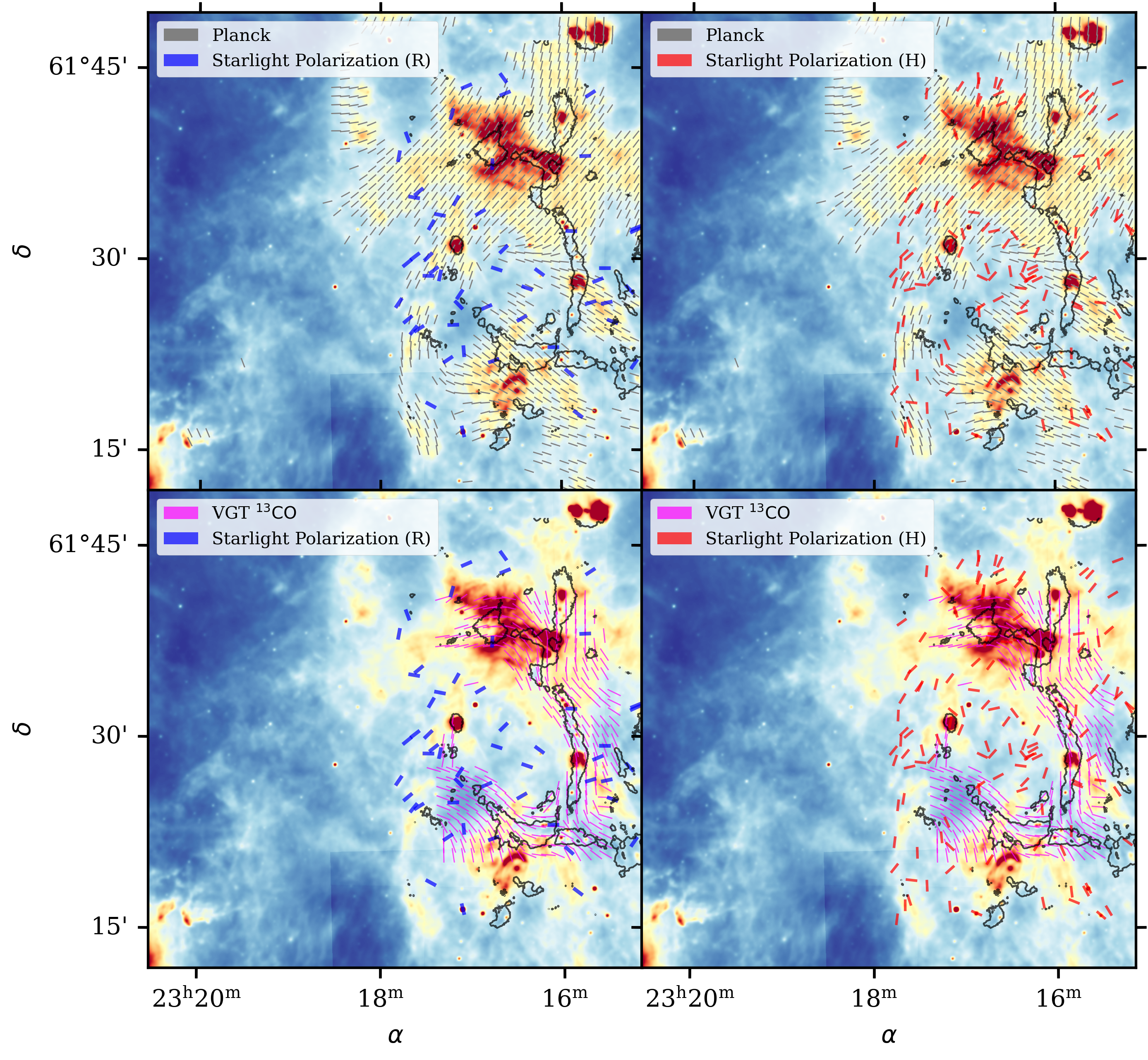}
    \caption{ WISE intensity maps of the {\cloudname} region with JCMT 0.2 mJy contours (black). The top panels show {\planck} polarization (gray) with starlight polarization in the R-band (blue, left) and H-band (red, right). The bottom panels display VGT polarization (magenta) with starlight polarization in the R-band (blue, left) and H-band (red, right).}
    \label{fig:wise}
\end{figure*}

We employ the WISE \citep{wright2010} Band 3 data to complement our analysis with the information on the distribution of the warm dust. 
Figure~\ref{fig:wise} presents the WISE $12\, \mu$m emission, overlaid with the magnetic field orientations derived from {\planck} sub-millimeter polarization , optical and near-infrared starlight polarization ({\bvis}, {\bnir}), and velocity gradients ({\bvgt}) from CO isotopologues.

The warm dust emission is more prominent in the eastern and southern parts of the cloud, where it overlaps with molecular structures traced by the CO emission.
The near-infrared compact sources align along this vertical cold dust filamentary ridge and form an arc. Another set of compact sources aligns along the inner front. 
The near-infrared map also shows an arc-like structure in the southern part which coincides with the structures detected in the JCMT map, pointing towards the south.

The magnetic field orientations show notable variations across different tracers. {\bplanck} exhibit large-scale uniformity. The rotation of {\bplanck} along and beyond the sub-millimeter cloud's elliptical ring is observed in the near-infrared, coinciding with the near-infrared intensity structures.
As already mentioned previously in Section~\ref{sec:b_polar}, foreground-subtracted visible and near-infrared polarization are in agreement with {\bplanck} in the southeastern part of {\cloudname}. In addition, {\bvis} and {\bnir} reveal finer-scale variations aligned with denser structures. {\bvgt} derived from {\coo} data follows the gas distribution closely, particularly in the northern and southern filamentary ridges.
 
These preliminary observations indicate a strong correlation between the magnetic field structure and the warm dust distribution, supporting the hypothesis that the cloud’s morphology is influenced by both magnetic and dynamical processes.
\revision{Conversely, the strong correlation may also indicate that the magnetic fields geometry is influenced by dynamic processes.}
Further interpretation of these findings is presented in Section~\ref{sec:disc_wise}.

\section{Discussion}
\subsection{Foreground magnetic field structure}
\label{sec:bforeground}
\begin{figure}
    \centering
    \includegraphics[width=1\linewidth]{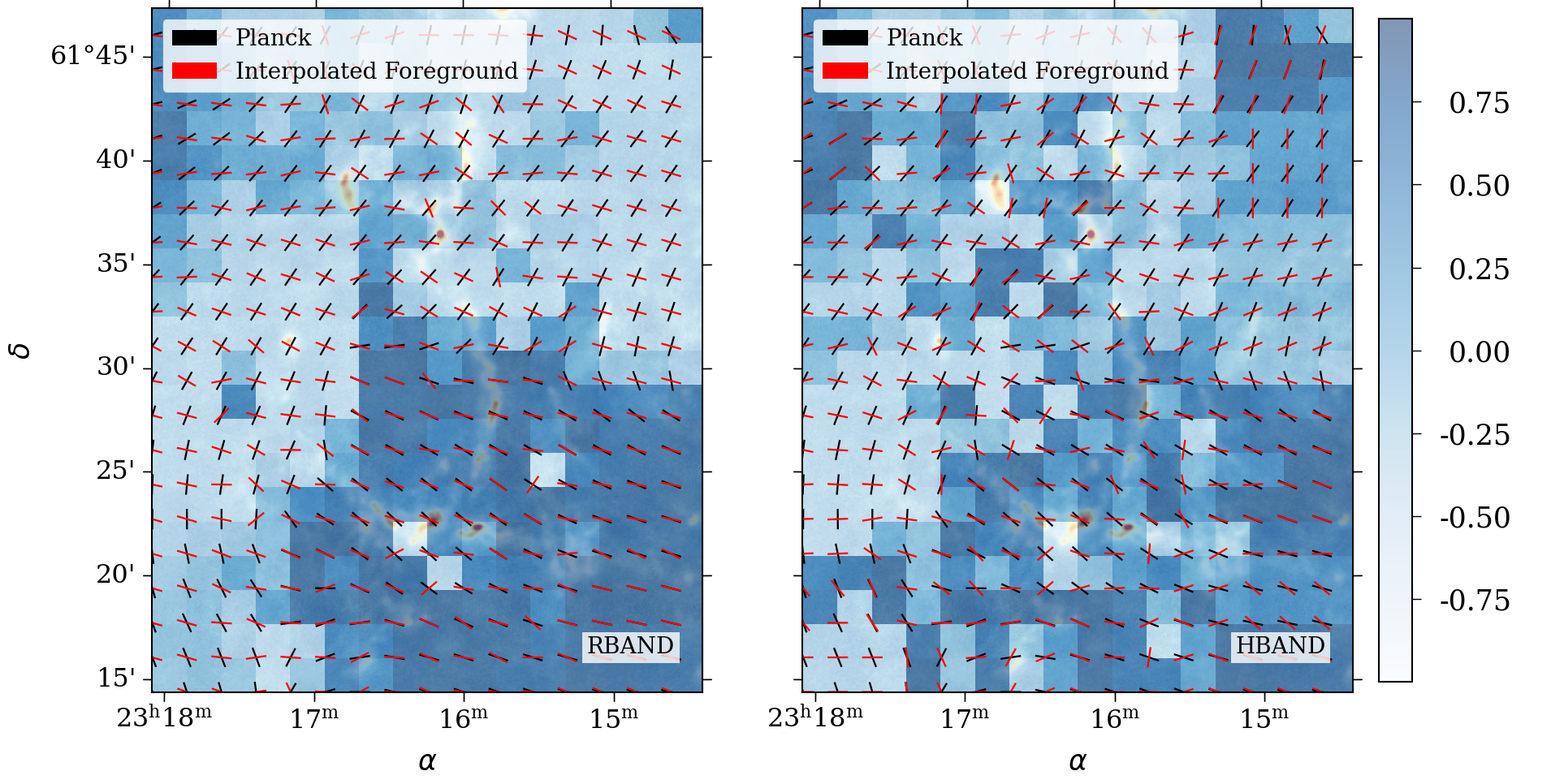}
    \caption{Alignment measure (AM) of the interpolated foreground magnetic field derived from the starlight polarization data for stars located at distances $[0;2500$] pc (green bi-vectors) and the magnetic field derived from 353 GHz (850 $\mu$m) {\planck} polarization data rotated by $90^{\circ}$ (black bi-vectors). \revision{AM = -1 means perpendicular relative orientation while AM = 1 means parallel relative orientation.} The background map is the JCMT intensity map at 850 $\mu$m as in Figure~\ref{fig:general}. Left panel: R band, right panel: H band.}
    \label{fig:vis-planck}
\end{figure}

The abrupt change in interstellar extinction (Figure~\ref{fig:avh-avr}) and Gaia magnitude (Figure \ref{fig:avh-avr-sc}) at 800 pc indicates that the increase in magnitude is due to the change in the dust distribution towards {\cloudname} and a significant amount of dust material in the foreground. 
In addition, the gradual increase of the lower envelope after that distance until the distance of the cloud's location, supports this hypothesis. 
In order to analyze the distant cloud's magnetic field structure, it is important to assess this foreground's magnetic field direction.
We compare the interpolated foreground {\bvis} and {\bnir} with {\bplanck}. Figure~\ref{fig:vis-planck} shows that in both bands, the polarization by extinction and the dust polarized emission traces the same interstellar dust regions in the southwestern to central parts, and in the northeastern part of the map. We associate this trend with the non-negligible foreground material strongly contributing to the sub-millimeter polarization. This is even more prominent in the R band, which does not probe very dense matter. \revision{The variability in the H band is probably due, first, to the sensitivity of the detector and second, to faint clouds that can be located along the line of sight.} 
The orientation of {\bplanck} and foreground {\bvis} observed in the southwestern part coincides with the Galactic plane orientation, which means that this orientation is the Galactic magnetic field, as was found out by \cite{hiltner1949} and later confirmed with the {\planck} observations \citep{planck2014-xix}.
\revision{Figure~\ref{fig:appendix_am_planck-starpol} in Appendix~\ref{sec:app}, where starlight polarization was smoothed to the {\planck} resolution, shows similar trends of the alignment measure.}
On the contrary, the diagonal SE-NW stripe of perpendicular alignment indicates a contribution of interstellar matter that is not traced by the foreground stars. In addition, we observe deviations of AM from those extrema at the position of the ring and the associated northern and southern structures.  

\subsection{{\bvgt} derived with different tracers}
\label{sec:disc_vgt}
The Velocity Gradients techniques allows us to probe the magnetic field structure at different densities by employing different tracers. We compare the results of {\bvgt} derived using {\coo} and {\cooo} observations with the help of Figure~\ref{fig:iram}.

Not surprisingly, {\coo}-{\bvgt} shows more features as it probes the broader structures of the molecular cloud complex, in addition to the filaments and cores.  
However, {\coo} becomes optically thick in very dense regions, so the {\cooo} complements the picture with the denser parts and allows us to probe the associated magnetic field.
In particular, we observe the difference between {\coo} and {\cooo}-derived {\bvgt} in the northern clump. This can be due to several reasons, such as the above-mentioned optical depth dependence \citep{hsieh2019}, dynamical effects such as gravitational pull or shocks, and physical differences of the regions with different densities within the cloud. In the latter case, the VGT can be seen as a powerful tool for the multi-scale tomography of the magnetic field, which was shown in the Vela-C molecular cloud's observational data \citep{hu2020}. We further investigate this region and the possible caveats of the VGT in Section~\ref{sec:vgtpol}.

\revision{There is} a global similarity between \revision{the TRAO {\co} (in Figure~\ref{fig:trao}) and the two IRAM,} {\coo} and {\cooo}, tracers. It indicates that the cloud is permeated by a magnetic field, which globally does not change \revision{from the envelope traced by {\co}, to the intermediate} density media traced by {\coo} and {\cooo} (approximately up to $10^3 - 10^4 \, \mathrm{cm}^{-3}$ according to \cite{evans1999}). 
It also indirectly indicates, according to the MHD-rooted basis of the VGT \citep{yuen2017vgt} that the gravity is not a dominant factor at the scales that we trace, which is about $\sim 0.5$ pc, taking into account the sub-block averaging. \revision{Gravitational effects would be observed as changes of the VGT-derived magnetic fields from the tracers of different density regimes.} Higher angular resolution observations in the sub-millimeter range would allow us to test the derived geometries and probe the magnetic field at smaller scales. 

\subsection{Comparison between {\bvis}, {\bnir}, and {\bvgt}}
\label{sec:vgtpol}
\begin{figure}
    \centering
    \includegraphics[width=1\linewidth]{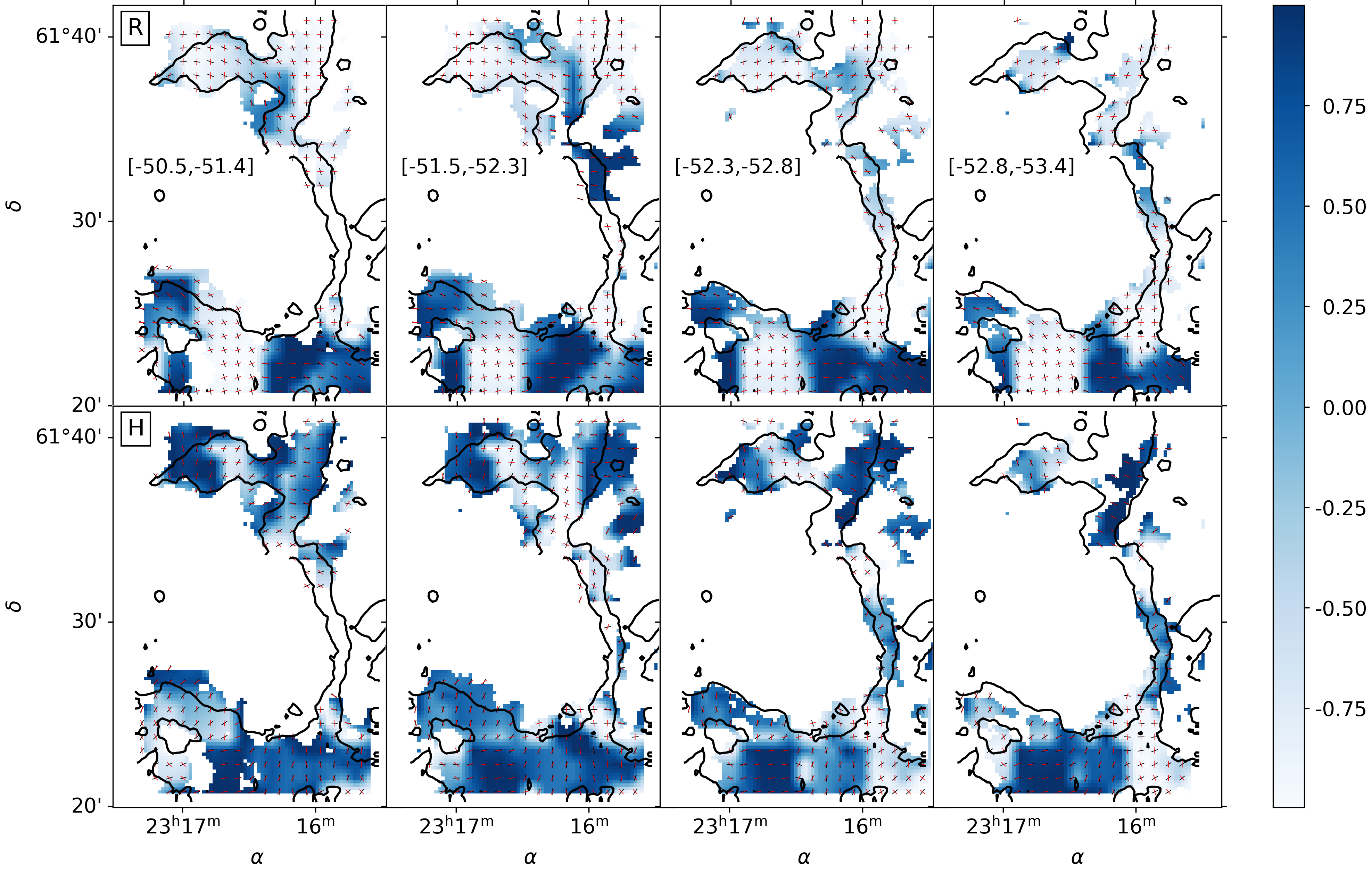}  
    \caption{Alignment measure (AM) between the magnetic field angles derived using foreground-subtracted starlight polarization (red segments) and using the velocity gradients technique (black segments) applied to {\coo} data. \revision{AM = -1 means perpendicular relative orientation while AM = 1 means parallel relative orientation.} Upper row: R band; lower row: H band. Each column corresponds to the different velocity ranges used for the {\bvgt} calculation. The contours correspond to the IRAM {\coo} contours at 15 K km/s as in Figure~\ref{fig:general}.}

    \label{fig:vgt-pol}
\end{figure}

We focus on {\coo} results as it traces matter at lower densities than {\cooo}, and hence can partly trace the near-infrared observations we compare. 
Although the optical polarization might not trace the same matter, we also include {\bvis} in the comparison.
Figure~\ref{fig:vgt-pol} shows the alignment measure AM between the magnetic field orientation derived using the foreground-subtracted starlight polarization data and the {\coo} velocity gradients technique. 

We note the most similarity between the magnetic field structures derived using the different methods in the channel [$-53.4;-52.5$] km/s, for both bands. This channel is centered on the part of the cloud that has the most prominent peak of emission, as discussed in \citet{frieswijk2007}. 
The alignment level is more significant between {\bvgt} and {\bnir} rather than {\bvis}, as expected  due to the near-infrared's ability to probe through denser parts of the cloud.

For the vertical filamentary ridge (\textit{e}) in the western part, we observe an agreement of the two angles in the H band,  with the magnetic field having a vertical orientation. In the R band, we observe only little agreement, which is probably motivated by the fact that observations at shorter wavelengths cannot probe dense material. The observed ridge, as we see in Figure~\ref{fig:iram} is either a single component, a filament, or a sheet, detected in the four velocity ranges with a uniform magnetic field orientation. 

In the southern part of the cloud, as noted in \cite{frieswijk2007} and \cite{fallscheer2013}, there are several velocity components. We observe a global agreement in the three velocity ranges that trace the southern half-ring (first, second, and third columns) \revision{for the structures \textit{b, d}, and \textit{g}}. Thus, the high AM indirectly suggests that the cloud's magnetic field has a curved structure in the southern part.

In the northern part of the cloud, there are both parallel and perpendicular relative orientations. However, in the northern clump (\textit{c}), there is a consensus on the parallel alignment between the polarization-derived B field and {\bvgt} except for the [$-53.4;-52.8$] km/s velocity range. 
The northern clump contains two compact sources, as seen in the thermal dust emission map in Figure~\ref{fig:general}. \cite{fallscheer2013} identified those cores as elongated structures with strong emission at longer \textit{Herschel} bands. 
The parallel alignment between the 90$^{\circ}$-rotated velocity gradients and polarization could be produced in case of a gravitational collapse, according to the findings by \cite{hu2020}. 
Thus, this region may undergo a gravitational collapse, which affects its velocity gradient and produces a change with respect to the large-scale component.
In addition, the discrepancy between the {\cooo}-derived and {\coo}-derived orientations (Figure~\ref{fig:iram}) in the range from -53.4 to -52.8 km/s favors this hypothesis.

\subsection{Comparison between {\bvis}, {\bnir}, and {\bplanck}}
\label{sec:b_polar}
\begin{figure}
    \centering
    \includegraphics[width=1\linewidth]{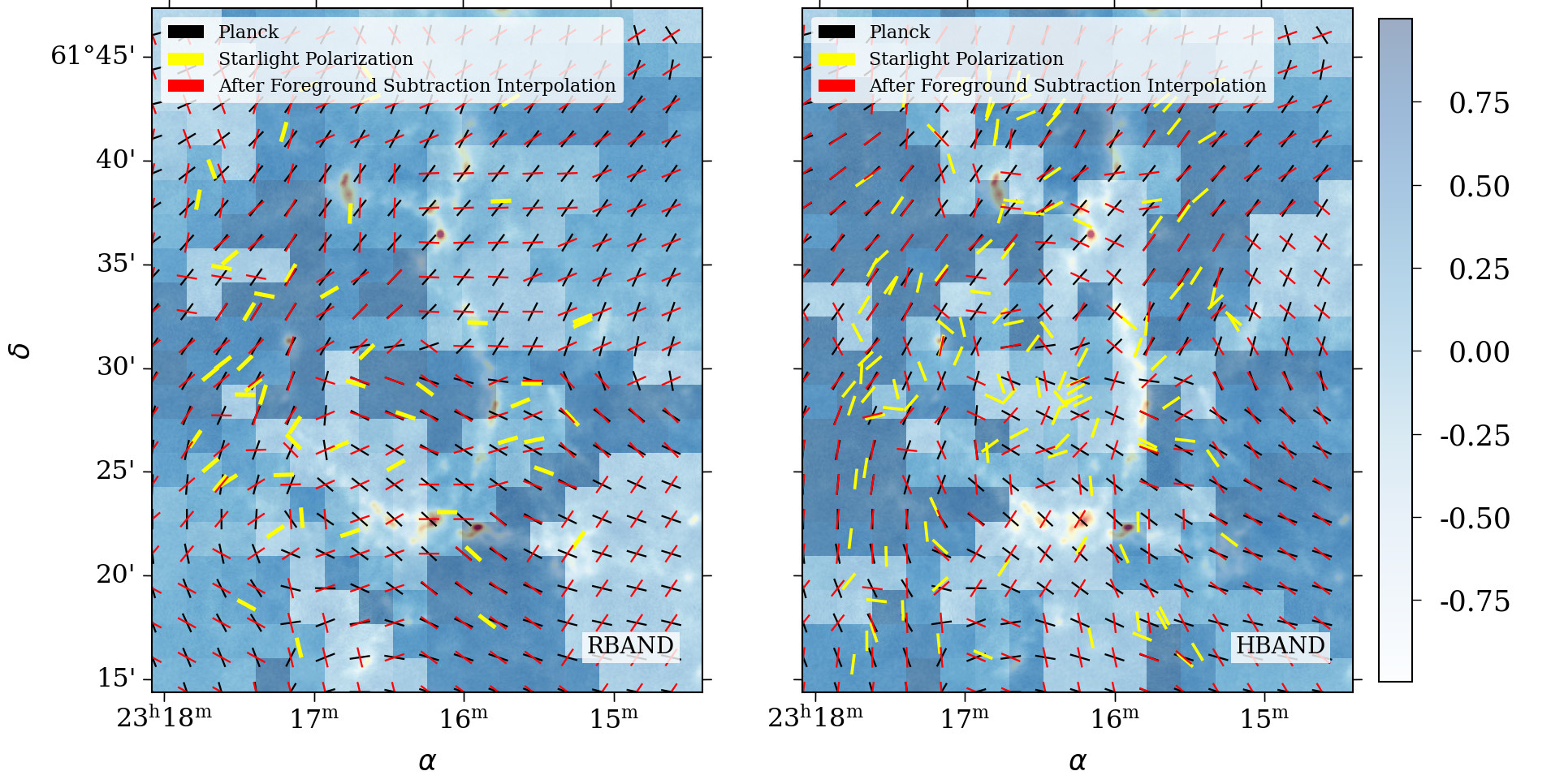}  
    \caption{ Alignment measure (AM) between the interpolated foreground-subtracted magnetic field derived from the starlight polarization data in {\cloudname} (black vectors) and the magnetic field derived from 353 GHz (850 $\mu$m) {\planck} polarization data. \revision{AM = -1 means perpendicular relative orientation while AM = 1 means paraller relative orientation.} The foreground-subtracted polarization angle towards the stars employed in the interpolation is shown in yellow. The background map is the JCMT intensity map at 850 $\mu$m as in Figure~\ref{fig:general}. Left
panel: R band, right panel: H band.}

    \label{fig:planck-starpol}
\end{figure}

We compare in Figure~\ref{fig:planck-starpol} the foreground-subtracted starlight polarization and the {\planck} polarization rotated by 90$^{\circ}$. 
Certainly, {\planck} data provides the line of sight integrated POS magnetic field. 
However, if most of the dust emission arises from a specific region along, the corresponding signal would be dominant. This is the case in the eastern part of the map, which showed no alignment when considering the foreground stars, shown in Figure~\ref{fig:fginterp}. On the contrary, the foreground subtracted starlight polarization is in agreement with the sub-millimetre data, in the eastern but also in the north-western parts of the map in Figure~\ref{fig:planck-starpol}. 
There, the alignment is better for the H band, which is expected due to the limited reach of the visible polarization signal in the R band. 

\subsection{Link to the region's dynamics}
\label{sec:disc_wise}

The WISE $12\, \mu$m map, presented in Figure~\ref{fig:wise}, reveals extended warm dust structures beyond the elliptical ring traced in the sub-millimeter data. These features likely result from stellar feedback, turbulent compression, or past shock interactions. Notably, the near-infrared emission aligns well with the magnetic field orientations derived from multiple tracers, providing further evidence of the interplay between dust dynamics and magnetic support.

In the eastern and southern parts of the cloud, where the $12\, \mu$m intensity is highest, we observe a close alignment between {\bvgt}, {\bvis}, and {\bnir}, suggesting that the gas and dust are dynamically coupled with the magnetic field. This region may have experienced external compression, leading to enhanced emission and structured magnetic alignment.

Conversely, in the northern clump, discrepancies between {\bvgt} and {\bvis}, and {\bnir} suggest that gravitational effects or shock-induced turbulence are modifying the local velocity structure. The perpendicular alignment between {\bvgt} derived from {\coo} and {\cooo} supports this interpretation, as such configurations are often associated with collapsing cores or compression fronts \citep{hu2020}.

The southern half-ring exhibits a particularly strong alignment between the inferred magnetic field orientation and the warm dust structure, indicating that magnetic tension may have played a role in shaping this feature. Additionally, the consistency between {\bplanck} and {\bvis} and {\bnir} in this region suggests that the observed field structure is influenced by an extended magnetic component, which could be either the large-scale Galactic magnetic field or the magnetic field of the original dense structures that were further influenced by compression.

These findings support the idea that the magnetic field contributes to the structural integrity of {\cloudname}, guiding the flow of gas and dust while interacting with external feedback mechanisms. High-resolution studies would be necessary to further test these interpretations and probe the small-scale magnetic field dynamics within the cloud and the link to star formation.

\subsection{G111 in the context of ring-like structures}
Ring-like structures are interesting targets for magnetic field observations as they allow testing the interplay between magnetic field support and molecular cloud formation. 
Studies like \cite{konyves2021} of the Rosette Nebula’s HII region and \cite{butterfield2024} of a large-scale ring-like structure using SOFIA HAWC+ data have demonstrated similar trends of alignment and compression. 
On one hand, \cite{konyves2021} studied a ring-like HII region within Rosette Nebula using the JCMT Bistro-2 Survey. The structure is associated with active star formation and its sub-millimeter polarization revealed a complex morphology of the magnetic field, which is partly aligned with the curved shape of the cloud, and is also partly perpendicular to it. In particular, parallel alignment was observed at the ionization front, while perpendicular alignment was observed for cold clumps. 
Despite the geometrical similarity, the current study and the study by \cite{konyves2021} differ in physical size and the origin of the ring-like structure. While \cite{konyves2021} analyzed a sub-parsec HII region, we focused on a several parsec scale molecular cloud that does not have a well-determined origin, and is a rather old star forming region.
On the other hand, \cite{butterfield2024} analyzed a ring-like structure of a size of {\cloudname}, that is of a several parsecs. The structure is seen in cold dust emission and the associated magnetic field was studied using the SOFIA HAWC+ data. The POS magnetic field was found to be aligned with the dust structures, suggesting, together with the available SiO and CS data \citep{tsuboi2015}, a shock-compression scenario. The estimated age is less than 2$\times 10^5$ years featuring an aged supernova remnant.   
Thus, {\cloudname}’s case remains unique due to its uncertain origin and older evolutionary stage. We observe that the filament-magnetic field alignment persisted throughout the evolution of the cloud.

\section{Conclusion and perspectives}
Our study aimed at tracing the magnetic field structure of the {\cloudname} molecular cloud with its distinct elliptical ring-like morphology within the hypothesis that the magnetic field configuration would provide us with a hint of the cloud's formation history and evolution.

We examined the plane of the sky magnetic field orientations at various scales and densities as well as the cold and warm ISM structures toward and within the cloud. We used a combination of sub-millimeter, optical, and molecular line observations.
We used the interstellar dust polarized emission from the {\planck} telescope to trace the large-scale magnetic field directions along the line of sight, while starlight extinction polarization from the Kanata telescope, combined with the Gaia star distance estimates allowed us to probe the foreground magnetic field structure toward {\cloudname} and to derive the cloud's magnetic field structure after foreground subtraction.
We also employed the velocity gradients technique, or the VGT, using CO isotopologues from IRAM 30m observations to investigate the magnetic field structure in denser regions of the cloud.

Our results reveal that {\cloudname} is permeated by a coherent yet spatially varying magnetic field, with variations in alignment across different tracers. The alignment between {\planck}-derived orientations and starlight polarization at distances smaller than 2500 pc indicated a significant foreground interstellar dust component, necessitating a careful foreground subtraction. Despite uncertainties in the Gaia-based distance estimates, the comparison of {\planck} and starlight polarization confirmed that {\planck} polarization in the southern and southwestern parts of the cloud arises from foreground interstellar dust, aligning with the Galactic magnetic field direction.

A strong agreement was found between starlight polarization for stars beyond the cloud's location and VGT-derived magnetic field orientations in the western ridge and southern parts of the cloud. This alignment suggests that the magnetic field follows the density structures and remains consistent across different densities in these regions. However, distinct orientations between CO isotopologues indicate that while the global field structure is coherent, local variations emerge in denser regions due to gravitational interactions and dynamics within the cloud. Furthermore, a significant correlation between the magnetic field structure and the warm dust distribution, particularly in the eastern and southern parts of the cloud, suggests the influence of stellar feedback.

\revision{We can relate our results to the groups of relative orientations outlined in Sect.~\ref{sec:intro} and conclude that the southern and eastern regions of G111 fall into the category of curved magnetic field configuration (group three), consistent with the magnetic draping or shock compression scenario. The western ridge presents an example of the parallel relative orientation scenario (group one). Perpendicular orientations are rare and are present in the northern region. Thus, the spatial variations of the relative orientations show the complex interplay between magnetic fields and the ISM.}

One of the main findings is the curved magnetic field observed along the dense ridges of {\cloudname} seen in the JCMT map, which is also mirrored in the WISE band 3 map, where the elliptical ring extends over a wider area. This alignment strongly suggests the influence of shock compression, contrasting with the perpendicular configuration expected in the absence of shocks. 
The orientation of magnetic fields along these dense structures supports that shock compression, likely from external forces such as stellar winds or supernova remnants, played a central role in shaping {\cloudname}. This is consistent with the theoretical framework of turbulent shock-driven compression such as by \citet{padoan2001} or \citet{federrath2016}. It can also be a result of shear flows that led to the formation and morphology of such clouds, proposed by \citet{hennebelle2013} or \citet{inoue2018}.

While the magnetic field may not be the sole cause of the elliptical shape of the cloud, it likely plays a role in maintaining and influencing its long-term evolution. In the absence of a sufficiently strong magnetic field, turbulent motions and thermal pressure could cause the structure to disperse more rapidly, or the cloud could fragment into smaller structures due to gravitational instability. A well-ordered magnetic field can help maintain coherence by guiding gas flows, resisting turbulent disruption, and regulating how material accumulates along filaments.

In conclusion, this work highlights the effectiveness of a multi-method approach to studying interstellar magnetic fields, demonstrating that integrating diverse techniques is essential for capturing the full scope of magnetic influences in complex cloud structures. Our findings reinforce the concept that shock-induced compression significantly shapes molecular cloud morphologies, with magnetic fields playing a crucial role in guiding gas flows and regulating star formation. Future high-resolution observations, in polarimetry and spectroscopy such as SiO and CS shock tracers, can provide critical insights for the advancement of our understanding of these processes within {\cloudname} and similar molecular clouds.

\appendix

\section{Alternative solution for the foreground contribution estimation}
\label{sec:app}

\begin{figure}
    \centering
    \includegraphics[width=0.7\linewidth]{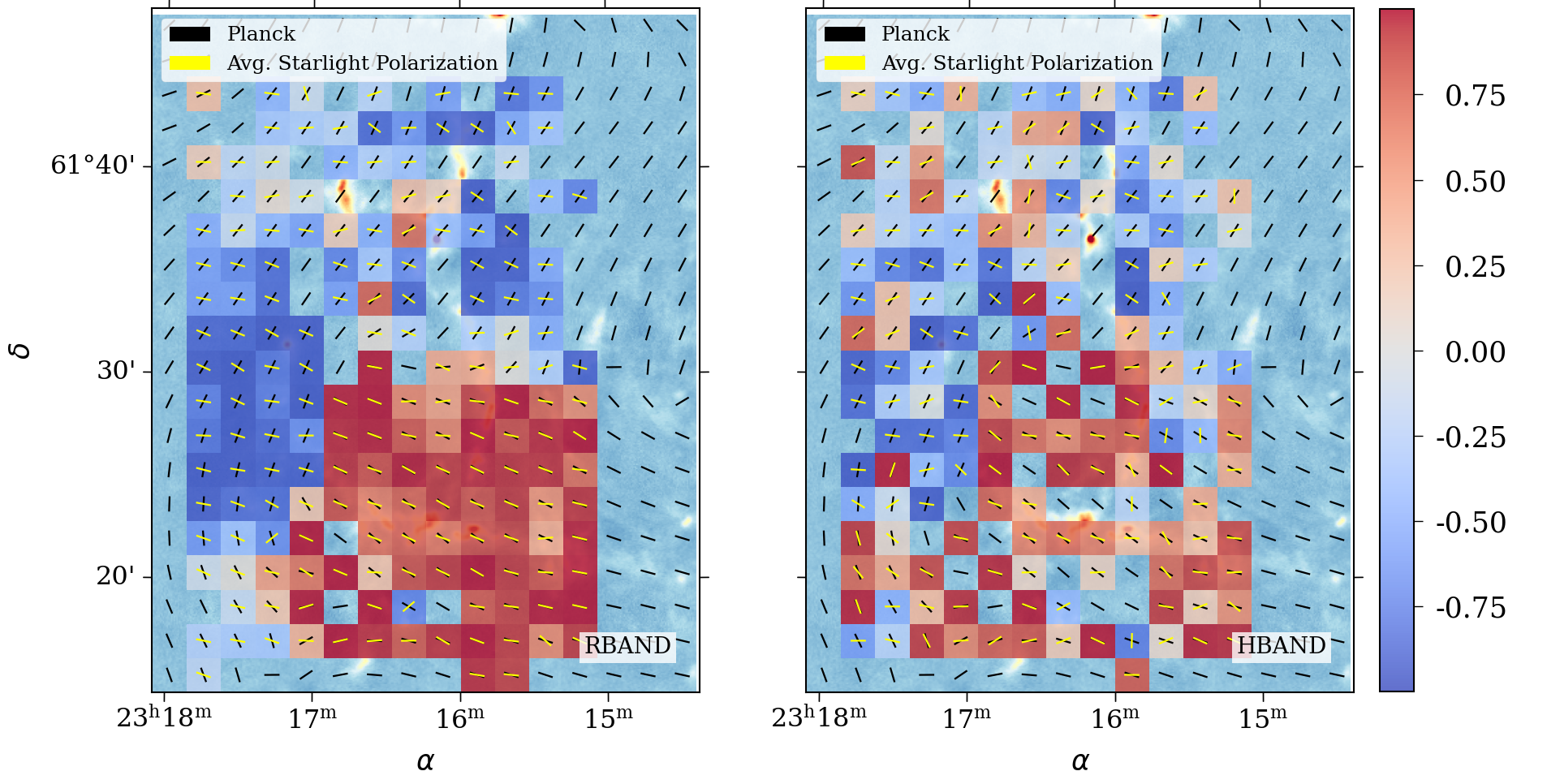}  
    \caption{ Alignment measure (AM) between Planck and starlight polarization orientations in the R-band (left) and H-band (right). }

    \label{fig:appendix_am_planck-starpol}
\end{figure}

\revision{To further investigate the consistency of the starlight polarization measurements interpolation, we smoothed the data to the {\planck} resolution. To do so, we averaged  polarization angles from starlight measurements within spatial bins matching the resolution of the {\planck} data. We computed the alignment measure (AM) between starlight and {\planck} polarizations similarly to Sect.~\ref{sec:bforeground} in both R and H bands. The resulting AM maps are shown in Figure~\ref{fig:appendix_am_planck-starpol}. Comparison to Figure~\ref{fig:planck-starpol} shows that both methods provide stable and consistent results.}

\section*{Acknowledgements}
This work was supported by the Faculty Development Competitive Research Grant Program of Nazarbayev University No. 201223FD8821. 
Y.H. acknowledges the support for this work provided by NASA through the NASA Hubble Fellowship grant \# HST-HF2-51557.001 awarded by the Space Telescope Science Institute, which is operated by the Association of Universities for Research in Astronomy, Incorporated, under NASA contract NAS5-26555.  A.L. acknowledges the support of NSF grants AST 2307840.
This work made use of Astropy:\footnote{http://www.astropy.org} a community-developed core Python package and an ecosystem of tools and resources for astronomy \citep{astropy:2013,astropy:2018,astropy2022}. We also acknowledge GILDAS \citep{gildas} software\footnote{https://www.iram.fr/IRAMFR/GILDAS} that was used to reduce the data.

\section*{Data Availability}
Data used in this study has been obtained as follows. The {\planck} data is publicly available at the {\planck} Legacy Archive, \url{https://pla.esac.esa.int/#home}; the WISE data is publicly available at the NASA/IPAC Insfrared Science Archive, \url{https://irsa.ipac.caltech.edu/frontpage/}; the IRAM 30m data is also publicly available through the IRAM Science Repository \url{https://iram-institute.org/science-portal/data-archive/}; the starlight polarization data can be provided upon reasonable request, subject to approval by the observatory director and of the principal investigator of the project.

\bibliography{thebibliography_all}{}

\begin{thebibliography}{}
\expandafter\ifx\csname natexlab\endcsname\relax\def\natexlab#1{#1}\fi
\providecommand{\url}[1]{\href{#1}{#1}}
\providecommand{\dodoi}[1]{doi:~\href{http://doi.org/#1}{\nolinkurl{#1}}}
\providecommand{\doeprint}[1]{\href{http://ascl.net/#1}{\nolinkurl{http://ascl.net/#1}}}
\providecommand{\doarXiv}[1]{\href{https://arxiv.org/abs/#1}{\nolinkurl{https://arxiv.org/abs/#1}}}

\bibitem[{D. {Abe} {et~al.}(2021){Abe}, {Inoue}, {Inutsuka}, \&
  {Matsumoto}}]{abe2021}
{Abe}, D., {Inoue}, T., {Inutsuka}, S.-i., \& {Matsumoto}, T. 2021,
  \bibinfo{title}{{Classification of Filament Formation Mechanisms in
  Magnetized Molecular Clouds},} \apj, 916, 83,
  \dodoi{10.3847/1538-4357/ac07a1}

\bibitem[{H. {Akitaya} {et~al.}(2014){Akitaya}, {Moritani}, {Ui}, {Urano},
  {Ohashi}, {Kawabata}, {Nakashima}, {Sasada}, {Sakimoto}, {Harao}, {Miyamoto},
  {Matsui}, {Itoh}, {Takaki}, {Ueno}, {Ohsugi}, {Nakaya}, {Yamashita}, \&
  {Yoshida}}]{akitaya2014}
{Akitaya}, H., {Moritani}, Y., {Ui}, T., {et~al.} 2014, \bibinfo{title}{{HONIR:
  an optical and near-infrared simultaneous imager, spectrograph, and
  polarimeter for the 1.5-m Kanata telescope},} in Society of Photo-Optical
  Instrumentation Engineers (SPIE) Conference Series, Vol. 9147, Ground-based
  and Airborne Instrumentation for Astronomy V, ed. S.~K. {Ramsay}, I.~S.
  {McLean}, \& H.~{Takami}, 91474O, \dodoi{10.1117/12.2054577}

\bibitem[{D. Alina {et~al.}(2019)Alina, Ristorcelli, Montier, Abdikamalov,
  Juvela, Ferri\`ere, Bernard, \& Micelotta}]{alina2019}
Alina, D., Ristorcelli, I., Montier, L., {et~al.} 2019,
  \bibinfo{title}{{Statistical analysis of the interplay between interstellar
  magnetic fields and filaments hosting Planck Galactic cold clumps},} \mnras,
  485, 2825, \dodoi{10.1093/mnras/stz508}

\bibitem[{D. {Alina} {et~al.}(2022){Alina}, {Montillaud}, {Hu}, {Lazarian},
  {Ristorcelli}, {Abdikamalov}, {Sagynbayeva}, {Juvela}, {Liu}, \&
  {Carri{\`e}re}}]{alina2021}
{Alina}, D., {Montillaud}, J., {Hu}, Y., {et~al.} 2022,
  \bibinfo{title}{{Large-scale magnetic field in the Monoceros OB 1 east
  molecular cloud},} \aap, 658, A90, \dodoi{10.1051/0004-6361/202039065}

\bibitem[{ {Astropy Collaboration} {et~al.}(2013){Astropy Collaboration},
  {Robitaille}, {Tollerud}, {Greenfield}, {Droettboom}, {Bray}, {Aldcroft},
  {Davis}, {Ginsburg}, {Price-Whelan}, {Kerzendorf}, {Conley}, {Crighton},
  {Barbary}, {Muna}, {Ferguson}, {Grollier}, {Parikh}, {Nair}, {Unther},
  {Deil}, {Woillez}, {Conseil}, {Kramer}, {Turner}, {Singer}, {Fox}, {Weaver},
  {Zabalza}, {Edwards}, {Azalee Bostroem}, {Burke}, {Casey}, {Crawford},
  {Dencheva}, {Ely}, {Jenness}, {Labrie}, {Lim}, {Pierfederici}, {Pontzen},
  {Ptak}, {Refsdal}, {Servillat}, \& {Streicher}}]{astropy:2013}
{Astropy Collaboration}, {Robitaille}, T.~P., {Tollerud}, E.~J., {et~al.} 2013,
  \bibinfo{title}{{Astropy: A community Python package for astronomy},} \aap,
  558, A33, \dodoi{10.1051/0004-6361/201322068}

\bibitem[{ {Astropy Collaboration} {et~al.}(2018){Astropy Collaboration},
  {Price-Whelan}, {Sip{\H{o}}cz}, {G{\"u}nther}, {Lim}, {Crawford}, {Conseil},
  {Shupe}, {Craig}, {Dencheva}, {Ginsburg}, {Vand erPlas}, {Bradley},
  {P{\'e}rez-Su{\'a}rez}, {de Val-Borro}, {Aldcroft}, {Cruz}, {Robitaille},
  {Tollerud}, {Ardelean}, {Babej}, {Bach}, {Bachetti}, {Bakanov}, {Bamford},
  {Barentsen}, {Barmby}, {Baumbach}, {Berry}, {Biscani}, {Boquien}, {Bostroem},
  {Bouma}, {Brammer}, {Bray}, {Breytenbach}, {Buddelmeijer}, {Burke},
  {Calderone}, {Cano Rodr{\'\i}guez}, {Cara}, {Cardoso}, {Cheedella}, {Copin},
  {Corrales}, {Crichton}, {D'Avella}, {Deil}, {Depagne}, {Dietrich}, {Donath},
  {Droettboom}, {Earl}, {Erben}, {Fabbro}, {Ferreira}, {Finethy}, {Fox},
  {Garrison}, {Gibbons}, {Goldstein}, {Gommers}, {Greco}, {Greenfield},
  {Groener}, {Grollier}, {Hagen}, {Hirst}, {Homeier}, {Horton}, {Hosseinzadeh},
  {Hu}, {Hunkeler}, {Ivezi{\'c}}, {Jain}, {Jenness}, {Kanarek}, {Kendrew},
  {Kern}, {Kerzendorf}, {Khvalko}, {King}, {Kirkby}, {Kulkarni}, {Kumar},
  {Lee}, {Lenz}, {Littlefair}, {Ma}, {Macleod}, {Mastropietro}, {McCully},
  {Montagnac}, {Morris}, {Mueller}, {Mumford}, {Muna}, {Murphy}, {Nelson},
  {Nguyen}, {Ninan}, {N{\"o}the}, {Ogaz}, {Oh}, {Parejko}, {Parley}, {Pascual},
  {Patil}, {Patil}, {Plunkett}, {Prochaska}, {Rastogi}, {Reddy Janga},
  {Sabater}, {Sakurikar}, {Seifert}, {Sherbert}, {Sherwood-Taylor}, {Shih},
  {Sick}, {Silbiger}, {Singanamalla}, {Singer}, {Sladen}, {Sooley},
  {Sornarajah}, {Streicher}, {Teuben}, {Thomas}, {Tremblay}, {Turner},
  {Terr{\'o}n}, {van Kerkwijk}, {de la Vega}, {Watkins}, {Weaver}, {Whitmore},
  {Woillez}, {Zabalza}, \& {Astropy Contributors}}]{astropy:2018}
{Astropy Collaboration}, {Price-Whelan}, A.~M., {Sip{\H{o}}cz}, B.~M., {et~al.}
  2018, \bibinfo{title}{{The Astropy Project: Building an Open-science Project
  and Status of the v2.0 Core Package},} \aj, 156, 123,
  \dodoi{10.3847/1538-3881/aabc4f}

\bibitem[{ {Astropy Collaboration} {et~al.}(2022){Astropy Collaboration},
  {Price-Whelan}, {Lim}, {Earl}, {Starkman}, {Bradley}, {Shupe}, {Patil},
  {Corrales}, {Brasseur}, {N{\"o}the}, {Donath}, {Tollerud}, {Morris},
  {Ginsburg}, {Vaher}, {Weaver}, {Tocknell}, {Jamieson}, {van Kerkwijk},
  {Robitaille}, {Merry}, {Bachetti}, {G{\"u}nther}, {Aldcroft},
  {Alvarado-Montes}, {Archibald}, {B{\'o}di}, {Bapat}, {Barentsen},
  {Baz{\'a}n}, {Biswas}, {Boquien}, {Burke}, {Cara}, {Cara}, {Conroy},
  {Conseil}, {Craig}, {Cross}, {Cruz}, {D'Eugenio}, {Dencheva}, {Devillepoix},
  {Dietrich}, {Eigenbrot}, {Erben}, {Ferreira}, {Foreman-Mackey}, {Fox},
  {Freij}, {Garg}, {Geda}, {Glattly}, {Gondhalekar}, {Gordon}, {Grant},
  {Greenfield}, {Groener}, {Guest}, {Gurovich}, {Handberg}, {Hart},
  {Hatfield-Dodds}, {Homeier}, {Hosseinzadeh}, {Jenness}, {Jones}, {Joseph},
  {Kalmbach}, {Karamehmetoglu}, {Ka{\l}uszy{\'n}ski}, {Kelley}, {Kern},
  {Kerzendorf}, {Koch}, {Kulumani}, {Lee}, {Ly}, {Ma}, {MacBride}, {Maljaars},
  {Muna}, {Murphy}, {Norman}, {O'Steen}, {Oman}, {Pacifici}, {Pascual},
  {Pascual-Granado}, {Patil}, {Perren}, {Pickering}, {Rastogi}, {Roulston},
  {Ryan}, {Rykoff}, {Sabater}, {Sakurikar}, {Salgado}, {Sanghi}, {Saunders},
  {Savchenko}, {Schwardt}, {Seifert-Eckert}, {Shih}, {Jain}, {Shukla}, {Sick},
  {Simpson}, {Singanamalla}, {Singer}, {Singhal}, {Sinha}, {Sip{\H{o}}cz},
  {Spitler}, {Stansby}, {Streicher}, {{\v{S}}umak}, {Swinbank}, {Taranu},
  {Tewary}, {Tremblay}, {de Val-Borro}, {Van Kooten}, {Vasovi{\'c}}, {Verma},
  {de Miranda Cardoso}, {Williams}, {Wilson}, {Winkel}, {Wood-Vasey}, {Xue},
  {Yoachim}, {Zhang}, {Zonca}, \& {Astropy Project Contributors}}]{astropy2022}
{Astropy Collaboration}, {Price-Whelan}, A.~M., {Lim}, P.~L., {et~al.} 2022,
  \bibinfo{title}{{The Astropy Project: Sustaining and Growing a
  Community-oriented Open-source Project and the Latest Major Release (v5.0) of
  the Core Package},} \apj, 935, 167, \dodoi{10.3847/1538-4357/ac7c74}

\bibitem[{C.~A.~L. {Bailer-Jones} {et~al.}(2021){Bailer-Jones}, {Rybizki},
  {Fouesneau}, {Demleitner}, \& {Andrae}}]{bailer-jones2021}
{Bailer-Jones}, C.~A.~L., {Rybizki}, J., {Fouesneau}, M., {Demleitner}, M., \&
  {Andrae}, R. 2021, \bibinfo{title}{{Estimating Distances from Parallaxes. V.
  Geometric and Photogeometric Distances to 1.47 Billion Stars in Gaia Early
  Data Release 3},} \aj, 161, 147, \dodoi{10.3847/1538-3881/abd806}

\bibitem[{S.~K. {Balfour} {et~al.}(2015){Balfour}, {Whitworth}, {Hubber}, \&
  {Jaffa}}]{balfour2015}
{Balfour}, S.~K., {Whitworth}, A.~P., {Hubber}, D.~A., \& {Jaffa}, S.~E. 2015,
  \bibinfo{title}{{Star formation triggered by cloud-cloud collisions},}
  \mnras, 453, 2471, \dodoi{10.1093/mnras/stv1772}

\bibitem[{L. {Barreto-Mota} {et~al.}(2021){Barreto-Mota}, {de Gouveia Dal
  Pino}, {Burkhart}, {Melioli}, {Santos-Lima}, \& {Kadowaki}}]{barretomota2021}
{Barreto-Mota}, L., {de Gouveia Dal Pino}, E.~M., {Burkhart}, B., {et~al.}
  2021, \bibinfo{title}{{Magnetic field orientation in self-gravitating
  turbulent molecular clouds},} \mnras, 503, 5425,
  \dodoi{10.1093/mnras/stab798}

\bibitem[{J.-P. {Bernard} {et~al.}(2023){Bernard}, {Bernard}, {Roussel},
  {Choubani}, {Alina}, {Aumont}, {Hughes}, {Ristorcelli}, {Stever},
  {Matsumura}, {Sugiyama}, {Komatsu}, {de Gasperis}, {Ferri{\`e}re}, {Guillet},
  {Ysard}, {Ade}, {de Bernardis}, {Bray}, {Crane}, {Dubois}, {Griffin},
  {Hargrave}, {Longval}, {Louvel}, {Maffei}, {Masi}, {Mot}, {Montel}, {Pajot},
  {P{\'e}rot}, {Ponthieu}, {Rodriguez}, {Sauvage}, {Savini}, {Tucker}, \&
  {Vacher}}]{bernard2023}
{Bernard}, J.-P., {Bernard}, A., {Roussel}, H., {et~al.} 2023,
  \bibinfo{title}{{Performance of the polarization leakage correction in the
  PILOT data},} Experimental Astronomy, 56, 197,
  \dodoi{10.1007/s10686-022-09882-5}

\bibitem[{J. {Brand} \& L. {Blitz}(1993){Brand} \& {Blitz}}]{brand1993}
{Brand}, J., \& {Blitz}, L. 1993, \bibinfo{title}{{The velocity field of the
  outer galaxy.},} \aap, 275, 67

\bibitem[{C.~M. {Brunt} {et~al.}(2003){Brunt}, {Kerton}, \&
  {Pomerleau}}]{brunt2003}
{Brunt}, C.~M., {Kerton}, C.~R., \& {Pomerleau}, C. 2003, \bibinfo{title}{{An
  Outer Galaxy Molecular Cloud Catalog},} \apjs, 144, 47,
  \dodoi{10.1086/344245}

\bibitem[{N.~O. {Butterfield} {et~al.}(2024){Butterfield}, {Guerra}, {Chuss},
  {Morris}, {Par{\'e}}, {Wollack}, {Costa}, {Hankins}, {Mackey}, {Staguhn}, \&
  {Zweibel}}]{butterfield2024}
{Butterfield}, N.~O., {Guerra}, J.~A., {Chuss}, D.~T., {et~al.} 2024,
  \bibinfo{title}{{SOFIA/HAWC+ Far-infrared Polarimetric Large Area CMZ
  Exploration Survey. II. Detection of a Magnetized Dust Ring in the Galactic
  Center},} \apj, 968, 63, \dodoi{10.3847/1538-4357/ad402c}

\bibitem[{D. {Clarke} {et~al.}(1983){Clarke}, {Stewart}, {Schwarz}, \&
  {Brooks}}]{clarke1983}
{Clarke}, D., {Stewart}, B.~G., {Schwarz}, H.~E., \& {Brooks}, A. 1983,
  \bibinfo{title}{{The statistical behaviour of normalized Stokes parameters},}
  \aap, 126, 260

\bibitem[{A. {Dolginov}(1972){Dolginov}}]{dolginov1972}
{Dolginov}, A. 1972, \bibinfo{title}{{Orientation of interstellar and
  interplanetary grains},} Astrophysics and Space Science, 18, 337,
  \dodoi{10.1007/BF00645399}

\bibitem[{B. {Draine} \& B. {Hansley}(2013){Draine} \& {Hansley}}]{draine2013}
{Draine}, B., \& {Hansley}, B. 2013, \bibinfo{title}{{MAGNETIC NANOPARTICLES IN
  THE INTERSTELLAR MEDIUM: EMISSION SPECTRUM AND POLARIZATION},} The
  Astrophysical Journal, 765, 159

\bibitem[{N.~J. {Evans}(1999){Evans}}]{evans1999}
{Evans}, II, N.~J. 1999, \bibinfo{title}{{Physical Conditions in Regions of
  Star Formation},} \araa, 37, 311, \dodoi{10.1146/annurev.astro.37.1.311}

\bibitem[{C. {Fallscheer} {et~al.}(2013){Fallscheer}, {Reid}, {Di Francesco},
  {Martin}, {Hill}, {Hennemann}, {Nguyen-Luong}, {Motte}, {Men'shchikov},
  {Andr{\'e}}, {Ward-Thompson}, {Griffin}, {Kirk}, {Konyves}, {Rygl},
  {Sadavoy}, {Sauvage}, {Schneider}, {Anderson}, {Benedettini}, {Bernard},
  {Bontemps}, {Ginsburg}, {Molinari}, {Polychroni}, {Rivera-Ingraham},
  {Roussel}, {Testi}, {White}, {Williams}, {Wilson}, {Wong}, \&
  {Zavagno}}]{fallscheer2013}
{Fallscheer}, C., {Reid}, M.~A., {Di Francesco}, J., {et~al.} 2013,
  \bibinfo{title}{{Herschel Reveals Massive Cold Clumps in NGC 7538},} \apj,
  773, 102, \dodoi{10.1088/0004-637X/773/2/102}

\bibitem[{C. {Federrath}(2016){Federrath}}]{federrath2016}
{Federrath}, C. 2016, \bibinfo{title}{{On the universality of interstellar
  filaments: theory meets simulations and observations},} \mnras, 457, 375,
  \dodoi{10.1093/mnras/stv2880}

\bibitem[{J. {Fenske} {et~al.}(2021){Fenske}, {Arakawa}, {Fallscheer}, \&
  {Francesco}}]{fenske2021}
{Fenske}, J., {Arakawa}, J., {Fallscheer}, C., \& {Francesco}, J.~D. 2021,
  \bibinfo{title}{{Runaway Stars as Possible Sources of the Elliptical Ring
  Structures in NGC 7538},} \aj, 161, 156, \dodoi{10.3847/1538-3881/abded2}

\bibitem[{M. {Fern{\'a}ndez-L{\'o}pez} {et~al.}(2021){Fern{\'a}ndez-L{\'o}pez},
  {Sanhueza}, {Zapata}, {Stephens}, {Hull}, {Zhang}, {Girart}, {Koch},
  {Cort{\'e}s}, {Silva}, {Tatematsu}, {Nakamura}, {Guzm{\'a}n}, {Nguyen Luong},
  {Guzm{\'a}n Ccolque}, {Tang}, \& {Chen}}]{fernandez-lopez2021}
{Fern{\'a}ndez-L{\'o}pez}, M., {Sanhueza}, P., {Zapata}, L.~A., {et~al.} 2021,
  \bibinfo{title}{{Magnetic Fields in Massive Star-forming Regions (MagMaR). I.
  Linear Polarized Imaging of the Ultracompact H II Region G5.89-0.39},} \apj,
  913, 29, \dodoi{10.3847/1538-4357/abf2b6}

\bibitem[{W.~W.~F. {Frieswijk} {et~al.}(2007){Frieswijk}, {Spaans}, {Shipman},
  {Teyssier}, \& {Hily-Blant}}]{frieswijk2007}
{Frieswijk}, W.~W.~F., {Spaans}, M., {Shipman}, R.~F., {Teyssier}, D., \&
  {Hily-Blant}, P. 2007, \bibinfo{title}{{Physical characteristics of a dark
  cloud in an early stage of star formation toward NGC 7538. An outer Galaxy
  infrared dark cloud?},} \aap, 475, 263, \dodoi{10.1051/0004-6361:20077148}

\bibitem[{ {Gildas Team}(2013){Gildas Team}}]{gildas}
{Gildas Team}. 2013, {GILDAS: Grenoble Image and Line Data Analysis Software},,
  Astrophysics Source Code Library, record ascl:1305.010

\bibitem[{P. {Goldreich} \& S. {Sridhar}(1995){Goldreich} \&
  {Sridhar}}]{goldreich1995}
{Goldreich}, P., \& {Sridhar}, S. 1995, \bibinfo{title}{{Toward a Theory of
  Interstellar Turbulence. II. Strong Alfvenic Turbulence},} \apj, 438, 763,
  \dodoi{10.1086/175121}

\bibitem[{D.~F. {Gonz{\'a}lez-Casanova} \& A.
  {Lazarian}(2017){Gonz{\'a}lez-Casanova} \& {Lazarian}}]{GL17}
{Gonz{\'a}lez-Casanova}, D.~F., \& {Lazarian}, A. 2017,
  \bibinfo{title}{{Velocity Gradients as a Tracer for Magnetic Fields},} \apj,
  835, 41

\bibitem[{F. {Heitsch}(2013){Heitsch}}]{heitsch2013}
{Heitsch}, F. 2013, \bibinfo{title}{{Gravitational Infall onto Molecular
  Filaments},} \apj, 769, 115, \dodoi{10.1088/0004-637X/769/2/115}

\bibitem[{P. {Hennebelle}(2013){Hennebelle}}]{hennebelle2013}
{Hennebelle}, P. 2013, \bibinfo{title}{{On the origin of non-self-gravitating
  filaments in the ISM},} \aap, 556, A153, \dodoi{10.1051/0004-6361/201321292}

\bibitem[{W. {Hiltner}(1949){Hiltner}}]{hiltner1949}
{Hiltner}, W. 1949, \bibinfo{title}{{On the Presence of Polarization in the
  Continuous Radiation of Stars. II.},} The Astrophysical Journal, 109, 471,
  \dodoi{10.1086/145151}

\bibitem[{T. {Hoang} {et~al.}(2018){Hoang}, {Cho}, \& {Lazarian}}]{hoang2018}
{Hoang}, T., {Cho}, J., \& {Lazarian}, A. 2018, \bibinfo{title}{{Alignment of
  Irregular Grains by Mechanical Torques},} \apj, 852, 129,
  \dodoi{10.3847/1538-4357/aa9edc}

\bibitem[{C.-h. {Hsieh} {et~al.}(2019){Hsieh}, {Hu}, {Lai}, {Yuen}, {Liu},
  {Hsieh}, {Ho}, \& {Lazarian}}]{hsieh2019}
{Hsieh}, C.-h., {Hu}, Y., {Lai}, S.-P., {et~al.} 2019, \bibinfo{title}{{Tracing
  Magnetic Field Morphology Using the Velocity Gradient Technique in the
  Presence of CO Self-absorption},} \apj, 873, 16,
  \dodoi{10.3847/1538-4357/ab0376}

\bibitem[{Y. {Hu} {et~al.}(2020){Hu}, {Lazarian}, \& {Yuen}}]{hu2020}
{Hu}, Y., {Lazarian}, A., \& {Yuen}, K.~H. 2020, \bibinfo{title}{{Velocity
  Gradient in the Presence of Self-Gravity: Identifying Gravity-induced Inflow
  and Determining Collapsing Stage},} arXiv e-prints, arXiv:2002.06754.
\newblock \doarXiv{2002.06754}

\bibitem[{Y. {Hu} {et~al.}(2018{\natexlab{a}}){Hu}, {Yuen}, \&
  {Lazarian}}]{hu2018}
{Hu}, Y., {Yuen}, K.~H., \& {Lazarian}, A. 2018{\natexlab{a}},
  \bibinfo{title}{{Improving the accuracy of magnetic field tracing by velocity
  gradients: principal component analysis},} \mnras, 480, 1333,
  \dodoi{10.1093/mnras/sty1807}

\bibitem[{Y. {Hu} {et~al.}(2018{\natexlab{b}}){Hu}, {Yuen}, \&
  {Lazarian}}]{HYL18}
{Hu}, Y., {Yuen}, K.~H., \& {Lazarian}, A. 2018{\natexlab{b}},
  \bibinfo{title}{{Improving the accuracy of magnetic field tracing by velocity
  gradients: principal component analysis},} \mnras, 480, 1333,
  \dodoi{10.1093/mnras/sty1807}

\bibitem[{Y. {Hu} {et~al.}(2019){Hu}, {Yuen}, {Lazarian}, {Ho}, {Benjamin},
  {Hill}, {Lockman}, {Goldsmith}, \& {Lazarian}}]{hu2019}
{Hu}, Y., {Yuen}, K.~H., {Lazarian}, V., {et~al.} 2019,
  \bibinfo{title}{{Magnetic field morphology in interstellar clouds with the
  velocity gradient technique},} Nature Astronomy, 3, 776,
  \dodoi{10.1038/s41550-019-0769-0}

\bibitem[{T. {Inoue} \& Y. {Fukui}(2013){Inoue} \& {Fukui}}]{inoue2013}
{Inoue}, T., \& {Fukui}, Y. 2013, \bibinfo{title}{{Formation of Massive
  Molecular Cloud Cores by Cloud-Cloud Collision},} \apjl, 774, L31,
  \dodoi{10.1088/2041-8205/774/2/L31}

\bibitem[{T. {Inoue} {et~al.}(2018){Inoue}, {Hennebelle}, {Fukui}, {Matsumoto},
  {Iwasaki}, \& {Inutsuka}}]{inoue2018}
{Inoue}, T., {Hennebelle}, P., {Fukui}, Y., {et~al.} 2018, \bibinfo{title}{{The
  formation of massive molecular filaments and massive stars triggered by a
  magnetohydrodynamic shock wave},} \pasj, 70, S53, \dodoi{10.1093/pasj/psx089}

\bibitem[{T. {Inoue} \& S.-i. {Inutsuka}(2012){Inoue} \&
  {Inutsuka}}]{inoue2012}
{Inoue}, T., \& {Inutsuka}, S.-i. 2012, \bibinfo{title}{{Formation of Turbulent
  and Magnetized Molecular Clouds via Accretion Flows of H I Clouds},} \apj,
  759, 35, \dodoi{10.1088/0004-637X/759/1/35}

\bibitem[{T. {Inoue} \& S.-i. {Inutsuka}(2016){Inoue} \&
  {Inutsuka}}]{inoue2016}
{Inoue}, T., \& {Inutsuka}, S.-i. 2016, \bibinfo{title}{{Formation of H I
  Clouds in Shock-compressed Interstellar Medium: Physical Origin of Angular
  Correlation between Filamentary Structure and Magnetic Field},} \apj, 833,
  10, \dodoi{10.3847/0004-637X/833/1/10}

\bibitem[{K.~S. {Kawabata} {et~al.}(1999){Kawabata}, {Okazaki}, {Akitaya},
  {Hirakata}, {Hirata}, {Ikeda}, {Kondoh}, {Masuda}, \& {Seki}}]{kawabata1999}
{Kawabata}, K.~S., {Okazaki}, A., {Akitaya}, H., {et~al.} 1999,
  \bibinfo{title}{{A New Spectropolarimeter at the Dodaira Observatory},}
  \pasp, 111, 898, \dodoi{10.1086/316387}

\bibitem[{V. {K{\"o}nyves} {et~al.}(2021){K{\"o}nyves}, {Ward-Thompson},
  {Pattle}, {Di Francesco}, {Arzoumanian}, {Chen}, {Diep}, {Eswaraiah},
  {Fanciullo}, {Furuya}, {Hoang}, {Hull}, {Hwang}, {Johnstone}, {Kang},
  {Karoly}, {Kirchschlager}, {Kirk}, {Koch}, {Kwon}, {Lee}, {Onaka},
  {Robitaille}, {Soam}, {Tahani}, {Tang}, {Tamura}, {Berry}, {Bastien},
  {Ching}, {Coud{\'e}}, {Kwon}, {Wang}, {Hasegawa}, {Lai}, \&
  {Qiu}}]{konyves2021}
{K{\"o}nyves}, V., {Ward-Thompson}, D., {Pattle}, K., {et~al.} 2021,
  \bibinfo{title}{{The JCMT BISTRO-2 Survey: The Magnetic Field in the Center
  of the Rosette Molecular Cloud},} \apj, 913, 57,
  \dodoi{10.3847/1538-4357/abf3ca}

\bibitem[{B.-C. {Koo} \& C.~F. {McKee}(1992){Koo} \& {McKee}}]{koo1992}
{Koo}, B.-C., \& {McKee}, C.~F. 1992, \bibinfo{title}{{Dynamics of Wind Bubbles
  and Superbubbles. I. Slow Winds and Fast Winds},} \apj, 388, 93,
  \dodoi{10.1086/171132}

\bibitem[{M. {Lachieze-Rey}(1981){Lachieze-Rey}}]{lachieze1981}
{Lachieze-Rey}, M. 1981, \bibinfo{title}{{Rayleigh Taylor Instabilities in the
  Interstellar Medium},} in Origin of Cosmic Rays, ed. G.~{Setti}, G.~{Spada},
  \& A.~W. {Wolfendale}, Vol.~94, 253

\bibitem[{A. Lazarian \& T. Hoang(2007)Lazarian \& Hoang}]{lazarian2007}
Lazarian, A., \& Hoang, T. 2007, \bibinfo{title}{{Subsonic mechanical alignment
  of irregular grains},} The Astrophysical Journal, 669, L77

\bibitem[{A. {Lazarian} \& D. {Pogosyan}(2000){Lazarian} \& {Pogosyan}}]{LP00}
{Lazarian}, A., \& {Pogosyan}, D. 2000, \bibinfo{title}{{Velocity Modification
  of H I Power Spectrum},} \apj, 537, 720, \dodoi{10.1086/309040}

\bibitem[{A. {Lazarian} \& E.~T. {Vishniac}(1999){Lazarian} \&
  {Vishniac}}]{LV99}
{Lazarian}, A., \& {Vishniac}, E.~T. 1999, \bibinfo{title}{{Reconnection in a
  Weakly Stochastic Field},} \apj, 517, 700, \dodoi{10.1086/307233}

\bibitem[{A. {Lazarian} \& K.~H. {Yuen}(2018{\natexlab{a}}){Lazarian} \&
  {Yuen}}]{lazarian2018vgt}
{Lazarian}, A., \& {Yuen}, K.~H. 2018{\natexlab{a}}, \bibinfo{title}{{Tracing
  Magnetic Fields with Spectroscopic Channel Maps},} \apj, 853, 96,
  \dodoi{10.3847/1538-4357/aaa241}

\bibitem[{A. {Lazarian} \& K.~H. {Yuen}(2018{\natexlab{b}}){Lazarian} \&
  {Yuen}}]{LY18a}
{Lazarian}, A., \& {Yuen}, K.~H. 2018{\natexlab{b}}, \bibinfo{title}{{Tracing
  Magnetic Fields with Spectroscopic Channel Maps},} \apj, 853, 96,
  \dodoi{10.3847/1538-4357/aaa241}

\bibitem[{H.-B. {Li} {et~al.}(2014){Li}, {Goodman}, {Sridharan}, {Houde}, {Li},
  {Novak}, \& {Tang}}]{li2014}
{Li}, H.-B., {Goodman}, A., {Sridharan}, T.~K., {et~al.} 2014,
  \bibinfo{title}{{The Link Between Magnetic Fields and Cloud/Star Formation},}
  Protostars and Planets VI, 101,
  \dodoi{10.2458/azu_uapress_9780816531240-ch005}

\bibitem[{T. {Liu} {et~al.}(2018){Liu}, {Kim}, {Juvela}, {Wang}, {Tatematsu},
  {Di Francesco}, {Liu}, {Wu}, {Thompson}, {Fuller}, {Eden}, {Li},
  {Ristorcelli}, {Kang}, {Lin}, {Johnstone}, {He}, {Koch}, {Sanhueza}, {Qin},
  {Zhang}, {Hirano}, {Goldsmith}, {Evans}, {White}, {Choi}, {Lee}, {Toth},
  {Mairs}, {Yi}, {Tang}, {Soam}, {Peretto}, {Samal}, {Fich}, {Parsons}, {Yuan},
  {Zhang}, {Malinen}, {Bendo}, {Rivera-Ingraham}, {Liu}, {Wouterloot}, {Li},
  {Qian}, {Rawlings}, {Rawlings}, {Feng}, {Aikawa}, {Akhter}, {Alina}, {Bell},
  {Bernard}, {Blain}, {B{\H{o}}gner}, {Bronfman}, {Byun}, {Chapman}, {Chen},
  {Chen}, {Chen}, {Chen}, {Chen}, {Chrysostomou}, {Cosentino}, {Cunningham},
  {Demyk}, {Drabek-Maunder}, {Doi}, {Eswaraiah}, {Falgarone}, {Feh{\'e}r},
  {Fraser}, {Friberg}, {Garay}, {Ge}, {Gear}, {Greaves}, {Guan},
  {Harvey-Smith}, {HASEGAWA}, {Hatchell}, {He}, {Henkel}, {Hirota}, {Holland},
  {Hughes}, {Jarken}, {Ji}, {Jimenez-Serra}, {Kang}, {Kawabata}, {Kim}, {Kim},
  {Kim}, {Kim}, {Koo}, {Kwon}, {Kuan}, {Lacaille}, {Lai}, {Lee}, {Lee}, {Lee},
  {Li}, {Li}, {Lo}, {Lopez}, {Lu}, {Lyo}, {Mardones}, {Marston}, {McGehee},
  {Meng}, {Montier}, {Montillaud}, {Moore}, {Morata}, {Moriarty-Schieven},
  {Ohashi}, {Pak}, {Park}, {Paladini}, {Pattle}, {Pech}, {Pelkonen}, {Qiu},
  {Ren}, {Richer}, {Saito}, {Sakai}, {Shang}, {Shinnaga}, {Stamatellos},
  {Tang}, {Traficante}, {Vastel}, {Viti}, {Walsh}, {Wang}, {Wang}, {Wang},
  {Ward-Thompson}, {Whitworth}, {Xu}, {Yang}, {Yang}, {Yuan}, {Zavagno},
  {Zhang}, {Zhang}, {Zhou}, {Zhou}, {Zhu}, {Zuo}, \& {Zhang}}]{liu2018b}
{Liu}, T., {Kim}, K.-T., {Juvela}, M., {et~al.} 2018, \bibinfo{title}{{The
  TOP-SCOPE Survey of Planck Galactic Cold Clumps: Survey Overview and Results
  of an Exemplar Source, PGCC G26.53+0.17},} \apjs, 234, 28,
  \dodoi{10.3847/1538-4365/aaa3dd}

\bibitem[{R. {Mazzei} {et~al.}(2023){Mazzei}, {Li}, {Chen}, {Fissel}, {Chen},
  \& {Park}}]{mazzei2023}
{Mazzei}, R., {Li}, Z.-Y., {Chen}, C.-Y., {et~al.} 2023,
  \bibinfo{title}{{Relative alignment between magnetic fields and molecular gas
  structure in molecular clouds},} \mnras, 521, 3830,
  \dodoi{10.1093/mnras/stad733}

\bibitem[{L. {Montier} {et~al.}(2015{\natexlab{a}}){Montier}, {Plaszczynski},
  {Levrier}, {Tristram}, {Alina}, {Ristorcelli}, \& {Bernard}}]{Montier1}
{Montier}, L., {Plaszczynski}, S., {Levrier}, F., {et~al.} 2015{\natexlab{a}},
  \bibinfo{title}{{Polarization measurements analysis. I. Impact of the full
  covariance matrix on the polarization fraction and angle measurements},}
  \aap, 574, A135

\bibitem[{L. {Montier} {et~al.}(2015{\natexlab{b}}){Montier}, {Plaszczynski},
  {Levrier}, {Tristram}, {Ristorcelli}, {Alina}, {Bernard}, \&
  {Guillet}}]{Montier2}
{Montier}, L., {Plaszczynski}, S., {Levrier}, F., {et~al.} 2015{\natexlab{b}},
  \bibinfo{title}{{Polarization measurements analysis. II. Best estimators of
  the polarisation fraction and angle},} \aap, 574, A136

\bibitem[{L. {Moscadelli} {et~al.}(2009){Moscadelli}, {Reid}, {Menten},
  {Brunthaler}, {Zheng}, \& {Xu}}]{moscadelli2009}
{Moscadelli}, L., {Reid}, M.~J., {Menten}, K.~M., {et~al.} 2009,
  \bibinfo{title}{{Trigonometric Parallaxes of Massive Star-Forming Regions.
  II. Cep A and NGC 7538},} \apj, 693, 406, \dodoi{10.1088/0004-637X/693/1/406}

\bibitem[{T. {Nagai} {et~al.}(1998){Nagai}, {Inutsuka}, \&
  {Miyama}}]{nagai1998}
{Nagai}, T., {Inutsuka}, S.-i., \& {Miyama}, S.~M. 1998, \bibinfo{title}{{An
  Origin of Filamentary Structure in Molecular Clouds},} \apj, 506, 306,
  \dodoi{10.1086/306249}

\bibitem[{J. {Naghizadeh-Khouei} \& D. {Clarke}(1993){Naghizadeh-Khouei} \&
  {Clarke}}]{naghizadeh-khouei1993}
{Naghizadeh-Khouei}, J., \& {Clarke}, D. 1993, \bibinfo{title}{{On the
  statistical behaviour of the position angle of linear polarization},} \aap,
  274, 968

\bibitem[{F. Nakamura \& Z.-Y. Li(2008)Nakamura \& Li}]{nakamura2008}
Nakamura, F., \& Li, Z.-Y. 2008, \bibinfo{title}{Magnetically regulated star
  formation in three dimensions: the case of the taurus molecular cloud
  complex,} \apj, 687, 354

\bibitem[{P. Padoan {et~al.}(2001)Padoan, Juvela, Goodman, \&
  Nordlund}]{padoan2001}
Padoan, P., Juvela, M., Goodman, A.~A., \& Nordlund, A. 2001,
  \bibinfo{title}{The Turbulent Shock Origin of Proto-Stellar Cores,} \apj,
  553, 227

\bibitem[{K. {Pattle} {et~al.}(2023){Pattle}, {Fissel}, {Tahani}, {Liu}, \&
  {Ntormousi}}]{pattle2023}
{Pattle}, K., {Fissel}, L., {Tahani}, M., {Liu}, T., \& {Ntormousi}, E. 2023,
  \bibinfo{title}{{Magnetic Fields in Star Formation: from Clouds to Cores},}
  in Astronomical Society of the Pacific Conference Series, Vol. 534,
  Protostars and Planets VII, ed. S.~{Inutsuka}, Y.~{Aikawa}, T.~{Muto},
  K.~{Tomida}, \& M.~{Tamura}, 193, \dodoi{10.48550/arXiv.2203.11179}

\bibitem[{ {Planck Collaboration} {et~al.}(2016){Planck Collaboration}, {Ade},
  {Aghanim}, {Alves}, {Arnaud}, {Arzoumanian}, {Ashdown}, {Aumont},
  {Baccigalupi}, {Banday}, {Barreiro}, {Bartolo}, {Battaner}, {Benabed},
  {Beno{\^\i}t}, {Benoit-L{\'e}vy}, {Bernard}, {Bersanelli}, {Bielewicz},
  {Bock}, {Bonavera}, {Bond}, {Borrill}, {Bouchet}, {Boulanger}, {Bracco},
  {Burigana}, {Calabrese}, {Cardoso}, {Catalano}, {Chiang}, {Christensen},
  {Colombo}, {Combet}, {Couchot}, {Crill}, {Curto}, {Cuttaia}, {Danese},
  {Davies}, {Davis}, {de Bernardis}, {de Rosa}, {de Zotti}, {Delabrouille},
  {Dickinson}, {Diego}, {Dole}, {Donzelli}, {Dor{\'e}}, {Douspis}, {Ducout},
  {Dupac}, {Efstathiou}, {Elsner}, {En{\ss}lin}, {Eriksen},
  {Falceta-Gon{\c{c}}alves}, {Falgarone}, {Ferri{\`e}re}, {Finelli}, {Forni},
  {Frailis}, {Fraisse}, {Franceschi}, {Frejsel}, {Galeotta}, {Galli}, {Ganga},
  {Ghosh}, {Giard}, {Gjerl{\o}w}, {Gonz{\'a}lez-Nuevo}, {G{\'o}rski},
  {Gregorio}, {Gruppuso}, {Gudmundsson}, {Guillet}, {Harrison}, {Helou},
  {Hennebelle}, {Henrot-Versill{\'e}}, {Hern{\'a}ndez-Monteagudo}, {Herranz},
  {Hildebrandt}, {Hivon}, {Holmes}, {Hornstrup}, {Huffenberger}, {Hurier},
  {Jaffe}, {Jaffe}, {Jones}, {Juvela}, {Keih{\"a}nen}, {Keskitalo}, {Kisner},
  {Knoche}, {Kunz}, {Kurki-Suonio}, {Lagache}, {Lamarre}, {Lasenby},
  {Lattanzi}, {Lawrence}, {Leonardi}, {Levrier}, {Liguori}, {Lilje},
  {Linden-V{\o}rnle}, {L{\'o}pez-Caniego}, {Lubin}, {Mac{\'\i}as-P{\'e}rez},
  {Maino}, {Mandolesi}, {Mangilli}, {Maris}, {Martin},
  {Mart{\'\i}nez-Gonz{\'a}lez}, {Masi}, {Matarrese}, {Melchiorri}, {Mendes},
  {Mennella}, {Migliaccio}, {Miville-Desch{\^e}nes}, {Moneti}, {Montier},
  {Morgante}, {Mortlock}, {Munshi}, {Murphy}, {Naselsky}, {Nati},
  {Netterfield}, {Noviello}, {Novikov}, {Novikov}, {Oppermann}, {Oxborrow},
  {Pagano}, {Pajot}, {Paladini}, {Paoletti}, {Pasian}, {Perotto}, {Pettorino},
  {Piacentini}, {Piat}, {Pierpaoli}, {Pietrobon}, {Plaszczynski},
  {Pointecouteau}, {Polenta}, {Ponthieu}, {Pratt}, {Prunet}, {Puget}, {Rachen},
  {Reinecke}, {Remazeilles}, {Renault}, {Renzi}, {Ristorcelli}, {Rocha},
  {Rossetti}, {Roudier}, {Rubi{\~n}o-Mart{\'\i}n}, {Rusholme}, {Sandri},
  {Santos}, {Savelainen}, {Savini}, {Scott}, {Soler}, {Stolyarov}, {Sudiwala},
  {Sutton}, {Suur-Uski}, {Sygnet}, {Tauber}, {Terenzi}, {Toffolatti}, {Tomasi},
  {Tristram}, {Tucci}, {Umana}, {Valenziano}, {Valiviita}, {Van Tent},
  {Vielva}, {Villa}, {Wade}, {Wandelt}, {Wehus}, {Ysard}, {Yvon}, \&
  {Zonca}}]{planck2016-XXXV}
{Planck Collaboration}, {Ade}, P.~A.~R., {Aghanim}, N., {et~al.} 2016,
  \bibinfo{title}{{Planck intermediate results. XXXV. Probing the role of the
  magnetic field in the formation of structure in molecular clouds},} \aap,
  586, A138, \dodoi{10.1051/0004-6361/201525896}

\bibitem[{ {Planck Collaboration Int. XIX}(2015){Planck Collaboration Int.
  XIX}}]{planck2014-xix}
{Planck Collaboration Int. XIX}. 2015, \bibinfo{title}{{\textit{Planck}
  intermediate results. XIX. An overview of the polarized thermal emission from
  Galactic dust},} \aap, 576, A104, \dodoi{10.1051/0004-6361/201424082}

\bibitem[{ {Planck Collaboration Int. XXXIII}(2016){Planck Collaboration Int.
  XXXIII}}]{planck2014-XXXIII}
{Planck Collaboration Int. XXXIII}. 2016, \bibinfo{title}{{\textit{Planck}
  intermediate results. XXXIII. Signature of the magnetic field geometry of
  interstellar filaments in dust polarization maps},} \aap, 586, A136

\bibitem[{E. {Purcell}(1979){Purcell}}]{purcell1979}
{Purcell}, E. 1979, \bibinfo{title}{{Suprathermal rotation of interstellar
  grains},} The Astrophysical Journal, 231, 404, \dodoi{10.1086/157204}

\bibitem[{G.~D. {Schmidt} {et~al.}(1992){Schmidt}, {Elston}, \&
  {Lupie}}]{schmidt1992}
{Schmidt}, G.~D., {Elston}, R., \& {Lupie}, O.~L. 1992, \bibinfo{title}{{The
  Hubble Space Telescope Northern-Hemisphere Grid of Stellar Polarimetric
  Standards},} \aj, 104, 1563, \dodoi{10.1086/116341}

\bibitem[{K. {Serkowski}(1958){Serkowski}}]{serkowski1958}
{Serkowski}, K. 1958, \bibinfo{title}{{Statistical analysis of the polarization
  and reddening of the double cluster in Perseus},} Acta Astronomica, 8, 135,
  \dodoi{10.1086/145465}

\bibitem[{J.~M. Stone {et~al.}(1998)Stone, Ostriker, \& Gammie}]{stone1998}
Stone, J.~M., Ostriker, E., \& Gammie, C. 1998, \bibinfo{title}{{Dissipation in
  compressible magnetohydrodynamic turbulence},} \apj, 508, L99

\bibitem[{J.~M. {Stone} \& E.~G. {Zweibel}(2009){Stone} \&
  {Zweibel}}]{stone2009}
{Stone}, J.~M., \& {Zweibel}, E.~G. 2009, \bibinfo{title}{{MHD Stability of
  Interstellar Medium Phase Transition Layers. I. Magnetic Field Orthogonal to
  Front},} \apj, 696, 233, \dodoi{10.1088/0004-637X/696/1/233}

\bibitem[{M. Tahani \&  et~al.(2023)Tahani \& et~al.}]{tahani2023}
Tahani, M., \& et~al. 2023, \bibinfo{title}{Magnetic Field Morphology around H
  II Regions: Tangential Fields and Implications for Star Formation,} ApJ, 944,
  139, \dodoi{10.3847/1538-4357/acac38}

\bibitem[{M. {Tsuboi} {et~al.}(2015){Tsuboi}, {Miyazaki}, \&
  {Uehara}}]{tsuboi2015}
{Tsuboi}, M., {Miyazaki}, A., \& {Uehara}, K. 2015,
  \bibinfo{title}{{Cloud-cloud collision in the Galactic center 50 km s$^{-1}$
  molecular cloud},} \pasj, 67, 109, \dodoi{10.1093/pasj/psv076}

\bibitem[{J. {Vaillancourt}(2007){Vaillancourt}}]{vaillancourt2007}
{Vaillancourt}, J. 2007, \bibinfo{title}{{Polarized Emission from Interstellar
  Dust},} EAS publication series, 23, 147

\bibitem[{E.~T. {Vishniac}(1994){Vishniac}}]{vishniac1994}
{Vishniac}, E.~T. 1994, \bibinfo{title}{{Nonlinear instabilities in
  shock-bounded slabs},} \apj, 428, 186, \dodoi{10.1086/174231}

\bibitem[{J. {Wardle} \& P. {Kronberg}(1974){Wardle} \&
  {Kronberg}}]{wardle1974}
{Wardle}, J., \& {Kronberg}, P. 1974, \bibinfo{title}{{The linear polarization
  of quasi-stellar radio sources at 3.71 and 11.1 centimeters},} \apj, 194, 249

\bibitem[{E.~L. {Wright} {et~al.}(2010){Wright}, {Eisenhardt}, {Mainzer},
  {Ressler}, {Cutri}, {Jarrett}, {Kirkpatrick}, {Padgett}, {McMillan},
  {Skrutskie}, {Stanford}, {Cohen}, {Walker}, {Mather}, {Leisawitz}, {Gautier},
  {McLean}, {Benford}, {Lonsdale}, {Blain}, {Mendez}, {Irace}, {Duval}, {Liu},
  {Royer}, {Heinrichsen}, {Howard}, {Shannon}, {Kendall}, {Walsh}, {Larsen},
  {Cardon}, {Schick}, {Schwalm}, {Abid}, {Fabinsky}, {Naes}, \&
  {Tsai}}]{wright2010}
{Wright}, E.~L., {Eisenhardt}, P. R.~M., {Mainzer}, A.~K., {et~al.} 2010,
  \bibinfo{title}{{The Wide-field Infrared Survey Explorer (WISE): Mission
  Description and Initial On-orbit Performance},} \aj, 140, 1868,
  \dodoi{10.1088/0004-6256/140/6/1868}

\bibitem[{S. {Xu} {et~al.}(2019){Xu}, {Ji}, \& {Lazarian}}]{xu2019}
{Xu}, S., {Ji}, S., \& {Lazarian}, A. 2019, \bibinfo{title}{{On the Formation
  of Density Filaments in the Turbulent Interstellar Medium},} \apj, 878, 157,
  \dodoi{10.3847/1538-4357/ab21be}

\bibitem[{K.~H. {Yuen} \& A. {Lazarian}(2017){Yuen} \&
  {Lazarian}}]{yuen2017vgt}
{Yuen}, K.~H., \& {Lazarian}, A. 2017, \bibinfo{title}{{Tracing Interstellar
  Magnetic Field Using Velocity Gradient Technique: Application to Atomic
  Hydrogen Data},} \apjl, 837, L24, \dodoi{10.3847/2041-8213/aa6255}

\end{thebibliography}
\bibliographystyle{aasjournalv7}

\end{document}